\documentclass[preprint]{elsarticle}
\usepackage{graphicx}
\usepackage{amsfonts}
\usepackage{amsmath}
\usepackage{amssymb}
\usepackage{amsthm}
\usepackage{subfigure}
\usepackage{a4wide}
\newcommand{\be}{\begin{equation}}
\newcommand{\ee}{\end{equation}}
\newcommand{\bel}[1]{\begin{equation}\label{#1}}
\newcommand{\bea}{\begin{eqnarray}}
\newcommand{\eea}{\end{eqnarray}}
\newcommand{\ba}{\begin{array}}
\newcommand{\ea}{\end{array}}

\newcommand{\mo}{\mathbb{E}}
\def\sign{\mbox{sign}}
\newtheorem{Proposition}{Proposition}

\begin{document}

\begin{center}
{\Large Short-time behaviour of demand and price viewed through an
exactly solvable model for heterogeneous interacting market agents}\\
\vskip3truemm
Gunter M. Sch\"utz$^{1,2}$, Fernando Pigeard de Almeida Prado$^{3}$, 
Rosemary J. Harris$^{4}$, Vladimir Belitsky$^{5}$
\end{center}
{\it (1) Institut f\"ur Festk\"orperforschung, Forschungszentrum
  J\"ulich, 52425 J\"ulich, Germany \\
(2) Interdisziplin\"ares Zentrum f\"ur Komplexe Systeme, Universit\"at Bonn, 
R\"omerstr. 164, 53117 Bonn, Germany\\
(3) Departamento de F\'isica e Matem\'atica, FFCLRP, Universidade de S\~ao Paulo,
14040-901, Ribeir\~ao Preto, SP, Brazil\\
(4) School of Mathematical Sciences,
Queen Mary University of London,
Mile End Road, London, E1 4NS, United Kingdom\\
(5) Instituto de Matem\'atica e Estat\'istica,
Universidade de S\~ao Paulo, Rua do Mat\~ao, 1010, CEP 05508-090,
S\~ao Paulo, SP, Brazil
}

\vspace{\baselineskip}

\noindent E-mail: {\tt g.schuetz@fz-juelich.de}

\begin{center}
June 16, 2009.
\end{center}

\noindent {\bf Abstract:}
We introduce a stochastic heterogeneous interacting-agent model for the
short-time non-equilibrium evolution of excess demand and price in a stylized 
asset market. We consider a combination of social interaction within peer groups
and individually heterogeneous fundamentalist trading decisions
which take into account the market price and the perceived fundamental 
value of the asset. The resulting excess demand is coupled to the market
price. Rigorous analysis reveals that this feedback may lead to price 
oscillations, a single bounce, or monotonic price behaviour. 
The model is a rare example of an analytically tractable interacting-agent model
which allows us to deduce in detail the origin of these different collective 
patterns. For a natural choice of initial distribution the results are 
independent of the graph structure that models the peer network of agents 
whose decisions influence each other.



\pagebreak


\section{Introduction}
\label{introd}

Heterogeneous Interacting-Agent Models (HIAMs) are stochastic processes
that are intensively employed to analyse and explain complex systems in 
diverse disciplines. In this paper, we propose a HIAM for analysing 
short-time fluctuations of an asset price in a single risky
asset market which is a stylized model of the real asset market. Even though
our model may describe quite diverse situations -- as any other HIAM does
-- we have chosen to discuss it in the framework of the asset market
because this setting facilitates the exposition and because of the broad interest
in understanding market behaviour. In this setting
the model explains the emergent behaviour of a large population of 
heterogeneous individual traders, where the behaviour of any single one is 
affected directly (not just indirectly via the asset price) by the behaviour of 
the whole population or a selected part of it.

We wish to point out that most HIAMs proposed in
the literature have been studied by numerical simulation or by some
heuristic approximation scheme and often these models are studied
primarily in the stationary regime. In contrast, both 
time-dependent and stationary properties of our model 
are tractable by exact analytical methods. This allows us to derive 
rigorously a number of non-trivial dynamic phenomena appearing in the
non-equilibrium collective behaviour of agents and
thus to obtain a detailed understanding of their origin. Before 
presenting (in Section 2) the
rigorous mathematical construction of our model,
we describe it informally and position it
in the world of HIAMs (which are extensively reviewed in
\cite{glaeser-scheinkman} and \cite{judd-tesfatsion}).

{\it Basic notions:} In our model, agents are market traders that at 
every time may behave in two ways: they either order to buy or order 
to sell a unit of a single asset (without knowing if their orders 
will be fulfilled or not). For simplicity of language, however, we
shall just say that an agent ``buys/sells'' a unit
of the asset, although we in fact mean ``orders to
buy/sell''. We shall also say that an agent is a buyer (seller) if her
current order is a buy-order (respectively, a sell-order).
According to the above, the excess demand of the asset is the
difference between the number of buyers and sellers. We use the term
\emph{relative excess demand} for the difference between buyers and
sellers divided by the total number of market agents (number of
buyers plus the number of sellers). 

A very important assumption of
the model is that (intermediate) transactions may occur out of
equilibrium, that is, when the excess demand is not zero. More
specifically, we assume that (intermediate) transactions are
permanently occurring in the market, independently of whether supply equals
demand or not. 

{\it Price evolution:} The price evolution of an asset in many regulated trading markets 
is determined by a series of rules known as the continuous double auction (see Ref.~\cite{Scalas} 
or Ref.~\cite{das}, the latter exposes these rules from a perspective linked to 
the ideas that form the basis of our model). Thus, when modeling this evolution, the principal 
question encountered is how precise these rules should be mirrored. Very faithful HIAMs of 
the  continuous double auction are not analytically tractable (see \cite{horst2} and references
therein). 
Being aware of this fact, we resorted to an alternative approach for mimicking price dynamics. 

There is a natural choice between two approaches to model price dynamics that ``dominate'' in 
the HIAM area.  They do not contradict the features that result from the rules of the  
continuous double auction; they just reflect this dynamics at different time scales 
mainly because they discard from consideration a part of those orders that are not 
fulfilled. Accordingly, by adopting these approaches, one can distort in a model some 
properties of the real flux of orders. A particularly common distortion is
the use of Poisson point processes: If a model would aim at mimicking faithfully the 
continuous double auction then it should incorporate that order arrival times do not form a 
Poissonian flow (see \cite{Scalas}). However, Poisson point process appear in almost all HIAMs 
because they underly the construction of continuous-time Markov processes 
which makes the model more amenable to analytical
analysis.. For this reason, also we use a Poisson point process in our model. In Ref.\cite{horst2} 
one finds some heuristics that suggests that a Poissonian flow is close to the flow of 
orders that are executed in the continuous double auction. 

One of the two dominating approaches mentioned above is grounded on the 
market-clearing condition which requires that the current excess demand is zero at the current 
asset price. Indeed, this may seem to be a reasonable criterion for the determination 
of the current asset price. Examples of works that have adopted this approach are
\cite{follmer-schweizer}, \cite{follmer-horst-kirman} and
\cite{horst}. However, since the current price is in fact a transaction price, the
market-clearing condition would exclude transactions occurring out
of equilibrium. 

In the second type of approach, the increment of 
the asset price (or its logarithm) increases with the relative
excess demand. Among the papers pioneering this approach (within the HIAM
setup) is \cite{beja-goldman}. As the title of that work
indicates, this approach is appropriate for studying a market out of
equilibrium. The principal reason for that is that the price
resulting from this rule does not enforce equality between the
numbers of the asset units bought and sold. The adoption of this approach
requires additional assumptions like the presence of a market maker
that adjusts the price, rather than the market-clearing condition 
as the determinant of the price.
In the present work we choose this second approach because it usually leads
to simpler stochastic and dynamical systems from which the price
process produces stylized trajectories. Moreover, our focus is on
the short-time market behavior where a market may indeed be far from
equilibrium.

We use the term ``price increment'' as a shorthand for the
increment between the current price and the price that
will be negotiated in the next transaction.  This price increment
will depend on the negotiation powers of sellers and
buyers.  A natural assumption is that the negotiation power of
sellers over buyers increases when the relative excess demand does
(the fewer sellers there are, the more power they have).
We account for this relationship by assuming that the current price return
is a linearly increasing function of the relative excess demand
(price return is the increment of the price logarithm). This price update
mechanism is already proposed by several authors, see e.g., \cite{lux},
\cite{kz1} and \cite{cmyz}. These considerations lead to the price update rule
\begin{equation}\label{mehir}\begin{array}{l}
\log\left(\hbox{price at time }t+dt\right)-
\log\left(\hbox{price at time }t\right) \equiv P(t+dt)- P(t)=\\
\kern12em=\lambda\, dt\, \times\left(\hbox{relative excess demand at
time } t \,
\right).\end{array}
\end{equation}
Here the positive model parameter
$\lambda$ is the feedback strength of the demand on the price. This
relationship reproduces the approximately linear part of the hyperbolic 
tangent relation reported in Ref. \cite{Pler02} for various stocks.

{\it Social interactions:} The interaction between agents in our model
is based on the notion 
that individual agents include in their judgement whether to buy or sell,
the decisions of peer traders, see e.g., \cite{kirman-ants} and \cite{Lux95}
for a brief survey and discussion. It is fair to assume that agents do not simply
follow the whole crowd. Instead, we postulate a herding mechanism
that is proportional to the buying (resp., selling) decisions only of
trusted peers of an agent. The strength of
this influence is quantified by a non-negative
parameter $\kappa$, called {\em the social
susceptibility}. The peer groups are modelled by a graph
structure: the agents are the nodes of a graph, and the
bonds pointing into a node define the peers of this node.
Notice that the individual structure of peer groups allows for modelling 
highly heterogeneous and directed social interaction, unlike
e.g., in the work \cite{Cont00} where undirected graphs
are considered.

It is worth noting that our theoretical result
(Proposition~\ref{expec}) describing the dynamics of the
expected relative excess demand, holds
for any graph of social interactions. This independence is an
interesting  property of our model. It adds information to the
understanding of how the topology of interactions influences the
generic properties of the HIAM. For example, if the graph topology is such that
the system (indexed by agents positions) is ergodic, then, by
evoking the system ergodicity, we can approach the dynamics of the
relative excess demand by the dynamics of its expected value
(described with Proposition~\ref{expec}). Our references for these
issues are \cite{liggett-social-network} and
\cite{topology-interaction}.

The agent interaction discussed above is just one of many
possible forms of social interaction generically called strategic
complementarity or positive externality and  thoroughly 
described in the reviews  \cite{glaeser-scheinkman} and
\cite{judd-tesfatsion}. The particular form reflected in our model
was chosen for the following two reasons. (1)~Since we intend
to model an asset market, we include a chartist behaviour: market
agents have the tendency to buy when the price increases and to sell
otherwise. Since in our model the price increment is coupled to the
relative excess demand, our interaction mechanism makes agents behave
in just this manner.
(2)~Our construction is also inspired by previous successful work that
used proportions of diverse market groups for modelling the formation
of strategies. This notion allows us to diversify the structure of
influences by attributing different peer groups to different
agents. This reflects market agents that try to draw information
about the best strategy (buy or sell) from the behaviour of their
peers. Accordingly, market agents will have the tendency to buy when
the majority of their peers buy and to sell otherwise. Of
particularly significant impact on this construction were the papers
\cite{lux} and \cite{bikhchandani}.

{\it Individual heterogeneity:} Another important characteristic of 
contemporary HIAMs
is the heterogeneity of agents. This means that agents behave
differently under similar circumstances. That this phenomenon is
present in real market agent behaviour has been substantiated by
firm arguments which are grounded on empirical facts and on economic
and psychological considerations, see e.g., \cite{hommes}. In our model agents 
are heterogeneous not only in the structure of their individual peer groups
but also in their evaluation of the fundamental value of the traded asset. This
individual heterogeneity is quantified by a probability distribution function
in the following way. Each time when an agent has to make up her mind
about the fundamental value, we sample $X_i(t)$, the ``noise'' in the evaluation
of agent $i$ at time $t$, from a mean-zero distribution, add an amount
$P_f(t)$ and
interpret this sum as the logarithm of the evaluation of the fundamental
value of this agent.  This implies that $P_f$ is the average -- taken over the whole population --
of the individual evaluations and that the cumulative distribution $\Phi$ of $X_i(t)$
has the following interpretation:
\begin{equation}\label{Phikak}\begin{array}{rcl}
\Phi(x)&=&\hbox{the proportion of the agents that think that the logarithm of}
\\
&&\hbox{the fundamental value is not greater than }P_f+x,\quad x\in \mathbb
{R}.\end{array}
\end{equation}
The behaviour of the model will depend on the function $P_f$, which for brevity 
we shall sometimes refer to as the ``average fundamental value evaluation''.
The form of $P_f(t)$ depends on all kinds of information which in the framework
of the efficient market hypothesis would be fully accounted for by the
market price $P(t)$, i.e., one would set $P_f(t) = P(t)$ if on average the 
agents believed in an efficient market. To allow for in principle arbitrary deviations 
from this idealized view we make a general ansatz
\begin{equation}\label{fund}
P_f(t)=U\left(P(t),t \right)
\end{equation}
where $U$ is a function which may be explicitly time-dependent
due to possible exogeneous influence that changes the average evaluation.
An appropriate choice of $U$ then
enables us to model a real asset market in which the average evaluation of the 
fundamental value
changes in time, caused by endogenous factors and/or by
exogenous factors.
Two simple
choices of this dependence are discussed in more detail in Sec. 4.
The distribution $\Phi$ does not change in time and, as a working hypothesis,
it appears reasonable to assume finite moments. In fact, we choose uniform i.i.d. 
noise  $X_i(t) \in [-\theta,\theta]$ so that the parameter $\theta$ is
a direct measure of the heterogeneity of the population of market agents.

We realize purely fundamentalist trading by the simplest, but still widely 
modelled, behaviour: to buy when the current
price is lower than the estimated fundamental value, and to sell
otherwise. Thus, in our model, agents have both speculative and fundamentalist
traits in their trading strategies. These components are combined
in the following way. An agent $i$ will buy (resp., sell) at time $t$,
if the following expression is positive (resp., negative):
\begin{equation}\label{bese}
\frac{\kappa}{\hbox{the number of the peers of }i}\sum_{j\,:\,
j\,\,\hbox{\scriptsize is a peer of }i}  s_j(t)
+ \Bigl(\left\{P_f(t)+X_i(t)\right\}-P(t)\Bigr).
\end{equation}
We shall refer to this expression as the {\it individual asset evaluation} (IAE).
Here, $s_j(t) \in \{-1,1\}$ denotes the position of the agent $j$
at time $t$: if $s_j(t) = 1$ ($s_j(t) = -1$), then the agent $j$ is
a buyer (seller) at time $t$.
Thus the second term of the IAE (\ref{bese}) corresponds to the
fundamentalist strategy described in the previous paragraph, where
$P_f(t)+X_i(t)$ stands for the logarithm of the fundamental value
evaluation of individual $i$ at time $t$. The first
term corresponds to the speculative strategy, because
$s_j(t)$ denotes the position ($1 =$ buys, $-1 =$ sells) of the
agent $j$ at time $t$. So the relation of the ``strengths'' of
these terms determines whether an agent acts more like
a speculator or like a fundamentalist.

This completes the informal description of our model. The main result (Proposition 1)
to be discussed and derived below contains the following: (1) Under some {\it average}
uniformity assumptions (detailed below) the expectation $\mo[s_i(t)]$
of the decision of agent $i$ at some time $t < t^\ast $ is the same for
{\it all agents} $i$ for {\it any} network topology.
The time $t^\ast$ up to which we can prove this result rigorously depends on
the choice of model parameters.
(2) For some natural choices of $P_f$ this expectation satisfies a linear
second order ODE (short for ordinary differential
equation) even though the model is intrinsically highly
non-linear. The statements (1) and (2) demonstrate that our model
provides a rare example of an non-trivial HIAM that is exactly solvable,
at least within some time interval $[0,t^\ast]$.
Numerical Monte Carlo simulations indicate an approximate validity of this
result for much longer times, but this is not really our concern, since we are
mainly interested in non-equilibrium short-time properties of the model.
More significantly, the simulations demonstrate that the collective patterns 
revealed by the
expectation $\mo[s_i(t)]$ are clearly recognizable in a single realization of the 
dynamics even though the process is noisy and the peer groups may be small
and highly heterogeneous.

These mathematical results are proved for two qualitatively different
assumptions on $P_f(t)$, viz.\ constant average fundamental value evaluation
(irrespective of the actual evolution of the asset price)
and an average fundamental value evaluation that follows the price up to
some systematic bias.
The ODE reveals interesting collective behaviour
of the agents since its solution may be either $i)$ a
monotonic function, or $ii)$ a function with unique extremal point,
or $iii)$ a damped oscillating function. From this we deduce under
which circumstances
the demand and the price
will be monotone in time, when they bounce once, and when they
oscillate. We give the conditions for each one of these three
possibilities in terms of the relation between the market parameters
that are reflected in our HIAM: the strength of social
susceptibility $\kappa$, the
heterogeneity of the fundamental value evaluations  parametrized by $\theta$,
and the strength of the demand feedback $\lambda$ and the initial
mismatch $\gamma$ between market price and average  value 
evaluation.

The paper is organized as follows. In Section~\ref{model}, we
present a precise definition of our HIAM and further comment on some
of its salient features. In order for this section to be
self-contained, we repeat some of the definitions already introduced above.
In Section~\ref{result}, we state the central result --
Proposition~\ref{expec}. In the main part of the paper, Section~\ref{application}, 
we elucidate some consequences of
Proposition~\ref{expec} to analyse short-time behaviour of demand and
price in an asset market. We also present the simulation results and 
discuss the role of the topology of peer groups. A concluding Section~\ref{Conclusions}
very briefly summarizes our main results and points to some
open questions and further related work. In the more technical Appendix~\ref{appendix}, we
present the proofs. This section can be skipped by a reader
interested in the results rather than their proofs.   
\footnote{Note that occasionally in
the paper we will draw on analogies with spin systems for the benefit
of readers familiar with the statistical mechanics literature.
However, we stress that no extensive use of physics connotations is
made and a reader without the relevant background can safely ignore
such references.}


\section{The Model}\label{model}

We start by establishing the structure of social influences among
agents. It is described by a finite directed graph that we denote by
$(\Lambda, \mathcal B)$. Each node $i \in \Lambda$ represents an agent
in the asset market. The cardinality $|\Lambda|$ of the set of nodes
is the total number of agents in the market. We put a bond $b_{ij}\in
\mathcal{B}$ pointing from $j$ to $i$ when we want to model that the
decision of the agent represented by node $i$ is influenced by
the opinion of the agent represented by node $j$. The set $J(i)$ of such nodes
forms the peer group of agent $i$ which we assume to be non-empty.
That is, our model has no ``lonely'' traders whose decisions are entirely
unaffected by their peers. All bonds in the graph
form the bond set $\mathcal{B}$. The actual structure of
$\mathcal{B}$ will not be specified now because it does not affect the validity of
our main result (Proposition~\ref{expec}).  Note that in the special
case where, for all pairs of nodes $i \in \Lambda, j \in \Lambda$, ``$j$ 
influences $i$'' $\Leftrightarrow$ ``$i$ influences $j$'', we have an 
undirected network.

For each node $i\in \Lambda$ and each time $t\geq 0$, we introduce the
binary variable $s_i(t)$ and call it {\em the state of $i$
at time $t$}.  Sometimes we will also refer to it as the ``position''
or ``decision'' of the agent -- as outlined earlier, we interpret $s_i(t)=+1$
(respectively, $-1$) as ``the agent $i$ wishes to buy (resp., sell)
one unit of stock at time $t$''. The quantity $s(t)$ denotes the set of all
$s_i(t)$ and  we postulate that $s_i(t)$ flips between $\pm1$  
as time $t\in [0,+\infty)$ goes on.

The time evolution of our model is a stochastic process $\{(s(t), P(t), P_f(t)),
t\geq 0\}$ with a family
of random variables $s(t)\in \{-1,1\}^\Lambda$, $P(t) \in \mathbb{R}$
(logarithm of the asset price), $P_f(t) \in \mathbb{R}$
(logarithm of the average evaluation of its fundamental value).
In slight abuse of language we shall, from now on, omit to state the
logarithmic relation between our model quantities and market
prices or values, and simply speak about price or fundamental value
respectively.

In order to define the stochastic dynamics we first recall that in order
to allow heterogeneity in the individual evaluations of the fundamental
value, we assume that $X_i(t_l^{(i)})$, $i \in \Lambda$, $l \in \{0,1,2
\ldots\}$ are independent and identically distributed random
variables, where for each $i$, $\{t_l^{(i)} \, :  \, l = 0,1,2 \ldots\}$ is an
arbitrary strictly increasing  sequence of times.
We remark that this use of annealed disorder (as opposed to
quenched disorder where $X_i$ would not change in time) is
appropriate for the absence of the market-clearing
condition: if a bid or ask of an agent has not been satisfied, she
may withdraw it and substitute it later by a bid or ask with another
price. For a discussion of quenched and annealed disorder see
\cite{gordon-seller} and \cite{nadal-multiple-equilibria}.
In our model with uniform distribution in $[-\theta,\theta]$
the individual evaluation $P_f(t) + X_i(t)$ of the fundamental value
at a given time $t$,
is a random variable from the set $[P_f(t)-\theta,P_f(t)+\theta]$.
The quantity $\theta$, proportional to the standard deviation of the
distribution, is a measure of the heterogeneity of the population
of agents.

Now we are in position to define the stochastic process $\{(s(t), P(t), P_f(t)),
t\geq 0\}$. Consider first the model parameter $P_f(t)$ which reflects what
the average of agents thinks what the fundamental value of the asset is.
Technically, the relation (\ref{fund}) in which we may choose the
parameter $P_f(t)$ to depend on the random variable $P(t)$ turns
$P_f(t)$ itself into a random variable even if $U$ is deterministic.
Two specific deterministic choices of the relationship $U$ will be
examined in detail in Section~\ref{application}. For such deterministic 
$U$ the stochastic dynamics
can be defined in terms of the random variables $s(t)$ and $P(t)$ alone.

It remains to define the stochastic evolution of $s(t)$ and $P(t)$.
Formally, the discrete decision process $s(t)$ in our model is the
first component of the stochastic process $\{(s(t), P(t), P_f(t)),
t\geq 0\}$. Given $P(t)$ we define the evolution of $s(t)$ using the Harris graphical 
approach, which is very popular in the Interacting Particle Systems field 
(see \cite{liggett}).  We employ the method here because it can be easily 
understood.

To each node $i\in \Lambda$ we attribute a Poisson point process
(PPP), that is, a set of marks $\tau_1^{(i)}, \tau_2^{(i)},\ldots$
on the ``time'' semi-line $[0,\infty)$ such that $\tau_1^{(i)},
\tau_2^{(i)}-\tau_1^{(i)}, \tau_3^{(i)}-\tau_2^{(i)}, \ldots$ are
independent random variables, each one having an exponential
distribution with mean $1$. The PPP's are postulated to be
independent across the population $\Lambda$. Each agent $i$, at each
mark $\tau_\ell^{(i)}$ of ``her'' PPP, draws a value $x^{(i)}_\ell$
from the random variable $X^{(i)}_\ell$ and sets her position at time
$t = \tau_\ell^{(i)}$ according to the following rule (see (\ref{bese}):
\begin{equation}\label{as}
s_i(t) = \left\{ \hbox{ \, is }\begin{array}{ll}+1,\hbox{ if }
\frac{\kappa}{|J(i)|} \sum_{j\in J(i)} s_j(t)+P_f(t) + x^{(i)}_\ell-P(t) \geq 0,\\
-1,\hbox{ if } \frac{\kappa}{|J(i)|} \sum_{j\in J(i)}
s_j(t)+P_f(t) + x^{(i)}_\ell-P(t) < 0. \end{array} \right.
\end{equation}
In terms of the
individual asset evaluation $\text{IAE}(i,t)$ (\ref{bese}), the decision rule takes
the simple intuitive
form
\begin{equation}\label{as2}
s_i(t) = \sign[\text{IAE}(i,t)].
\end{equation}
In order to define $s_i(t)$ for $t \notin \{\tau_1^{(i)},
\tau_2^{(i)}, \ldots\}$, we assume that $s_i(t)$ does not vary
during two subsequent decision times, that is, $s_i(t) =
s_i(\tau_\ell)$  for $t \in [\tau_\ell, \tau_{\ell+1})$, $\ell =
1,2,\ldots$.
The independence of the PPPs ensures that the probability of two or
more marks to coincide is zero and hence, for fixed $P(t)$ and $P_f(t)$, the decision rule
(\ref{as}) and the preceding assumption  define uniquely  the value of $s_i(t)$
(except on a measurable set of probability zero). Notice that we assume a PPP in order to keep the
model simple and analytically tractable. For a discussion of non-exponential
interarrival times of decisions see, e.g., \cite{Scalas}.

Using (\ref{mehir}) we can now define the evolution of $P(t)$ as
\begin{equation}\label{evp}
P(t)=P(0)+\frac{\lambda}{|\Lambda|}\int_0^t \sum_{i\in
\Lambda}s_i(u)du.
\end{equation}
This  is the time-integrated form of the widely used assumption  (\ref{mehir}) that
the price change at time $t$ is proportional to the excess demand at that instant of time,
which we discussed in the introduction.
In the previous graphical construction the value of $P(t)$ entering (\ref{as})
is understood as the price immediately before the decision taken at time
$t$.

We remind the reader that the determination of $P(t)$, according
to (\ref{evp}), does \emph{not} result from the market-clearing condition
which would instead require that the current excess
demand is zero at the current asset price.
Although, in general, the market-clearing condition must be satisfied
at equilibrium prices, it excludes an important factor which is
crucial for  price dynamics in speculative markets. Namely, that
some transactions occur -- and consequently transaction prices are
registered -- out of equilibrium, that is, at prices where demand and
supply do not match. Since in speculative markets, demand and supply
are strongly influenced by the observed price history, this
apparently harmless fact leads to completely different price
dynamics to those resulting from a market-clearing condition -- see
\cite{takayama} for an interesting discussion.
In fact, it can be shown that the adoption of a market-clearing
condition in our model would make our asset price fluctuate
around $P_f$ (the average evaluation of the fundamental value).
This would exclude stylized facts like price bubbles ($P(t)$ moving
away from $P_f$) and price crashes.
In order to study the effect of transactions occurring out of
equilibrium on price dynamics, we assume that the observed
transaction prices change little by little in accordance with the
pressure of non-zero excess demands divided by the trading volume.
The simplest way to model this dependence is given by (\ref{evp}).
Similar price mechanisms are also considered in \cite{lux}, \cite{kz1}
and \cite{cmyz}.

The process $\left\{\left(s(t), P(t), P_f(t))\right),\, t\geq 0\right\}$ that we have just
constructed is Markov. This is worth noting because the markovness is essentially 
employed in the proof of our main result (Proposition~1). This property is yielded by 
the very construction. Let us be more specific in respect.  
The first cornerstone that supports the markovness is the lack of memory property of 
each one of the Poisson point processes that we use to determine the times of 
position updates of the agents. The second cornerstone is the rule (\ref{as}) 
that is tailed so to ensure that an update of an agent's position,
when its time comes, depends solely on the state of the process's entities at the time 
of update (namely, $s(\cdot), P(\cdot)$and $P_f(\cdot)$) and of the entities that are independent
of everything else (namely, $X_\cdot(\cdot)$). 
The third cornerstone for the markovness of our process is the following fact: given the history 
of the process on some time interval $[0,t]$, the distribution of $P(s)$ at some given $s>t$
depends solely on the process' state at $t$; this fact follows readily from eq.~(\ref{evp}) since 
it can be re-written as $P(s)=P(t)+\lambda |\Lambda|^{-1}\int_t^s\sum_{i\in \Lambda}s_i(u)du$ 
using the additivity of the integral. The same fact holds true for $P_f(\cdot)$, but now it is 
a direct consequence of our definition (3). This 
is the fourth and the last cornerstone that supports the markovness of the process  
 $\left\{\left(s(t), P(t), P_f(t))\right),\, t\geq 0\right\}$. We shall only add in respect that
none of the components of these triple process -- if taken separately -- would be a Markov process.  


\section{Exact results}
\label{result}

The result to be formulated in the present section applies to our
model when the initial distribution of the node states satisfies a
particular property. This property is formulated in the 
paragraph below, and we shall also argue that it naturally holds
for a wide class of real agent market position distributions.

We say that a random $\{-1,1\}^\Lambda$-valued element $s$ has  {\em
network homogeneous mean}, if the expectation $\mo[s_i]$ is the same
for each $i\in \Lambda$ (recall, that $s_i$ denotes the value of $s$ at
the site $i\in \Lambda$). The simplest example of a
network homogeneous mean distribution is the product measure on
$\{-1,1\}^\Lambda$ that assigns the same probability, $p$ say,  to
$\{s_i=+1\}$ for each $i\in \Lambda$. We believe that one would
take this product measure for the initial node state distribution, if one
wished to express that initial decisions of the agents in a market are similar
and independent.

Note, however, that network homogeneous mean does not
require that the random variables $s_i$, $i\in \Lambda$, are
stochastically independent from each other. As a simple
counterexample, consider an Ising model
on the torus in $\mathbb{Z}^d$ (for some $d\in \mathbb{N}$) with a
non-vanishing
nearest neighbour potential $\{J\}$ and an external field $h$. 
Its equilibrium distribution, the Gibbs measure, is known to have correlated spins
and, as it may be shown easily, has net homogeneous mean in the sense of our
definition.

We have so far introduced all the concepts necessary for the
formulation of our main result (Proposition~\ref{expec} below). 
At this point we remind the reader of the meaning of the working 
parameters of the model and its output quantities:
\begin{description}
\item{$s_i(t)$} is the state of node $i\in \Lambda$ at time $t$. 
We say that the node $i$ buys [resp., sells] one unit of the asset
at time $t$, if $s_i(t)=+1$ [resp., $s_i(t)=-1$]. The quantity
$s(t) = \{s_i(t)\,:\, i \in \Lambda\}$ denotes
the whole configuration of states at time $t$. The evolution is defined in (\ref{as}).
\item{$P(t)$} denotes the price of the asset at time $t$. The evolution is defined in (\ref{evp}).
\item{$P_f(t)$} denotes the average evaluation of the fundamental value of the asset at time $t$.
The evolution is defined by the function $U$ in (\ref{fund}) and is further 
specified below.
\item{$\theta$} is proportional to the standard deviation of the heterogeneity of agents,
i.e., of the deviations from $P_f(t)$ of the individual evaluations of the fundamental value.
\item{$\kappa$} is the strength of social susceptibility and means
the non-negative weight with which the decision of an agent is influenced
by her peer group.
\item{$\lambda$} is the strength of the demand feedback on the price.
\end{description}

We now state the main result for our model (see the appendix for the proof).

\begin{Proposition}\label{expec}
Consider the continuous-time stochastic process $\{(s(t), P(t), P_f(t)),
t\geq 0\}$ defined in Section~\ref{model}, on an arbitrary  finite
oriented graph where each node has
at least one peer. Suppose the model satisfies the
following conditions:
\begin{description}
\item{(i)} the distribution of  the initial node states $s(0)$ is a
network homogeneous mean distribution;
\item{(ii)} $\Phi$ is the cumulative distribution with uniform density on
$[-\theta,+\theta]$ for some $\theta > \kappa$ (recall that $\kappa
\geq 0$).
\end{description}

Assume there exists $t^\ast$ such that 
\bel{tast} |P(t)-P_f(t)|\leq
\theta-\kappa \hbox{ for all }t\in [0,t^\ast].
\ee

Then, 
\begin{description}
\item{(a)} for every $t\in[0, t^\ast]$ and for each $i\in \Lambda$, it holds that
$\mo[s_i(t)]=f(t)$, where $f(t)$ is the solution of the following
integral-differential equation:
\begin{equation}\label{difeq}
\frac{d}{dt}f(t)=-\left(1-\frac{\kappa}{\theta}\right)f(t)
-\frac{\lambda}{\theta}\int_0^t f(u)du-\frac{1}{\theta}P(0)
+\frac{1}{\theta}\mo\left[P_f(t)\right]
\end{equation}
with the initial condition
\begin{equation}\label{inco}
f(0)=\mo[s_i(0)],
\end{equation}
\item{(b)} and, in particular, the distribution of the node states $s(t)$
has network homogeneous mean.
\end{description}
\end{Proposition}

Phrased in statistical mechanics language we solve with
Proposition \ref{expec} the evolution for a class
of stochastic Ising-type processes in which the external field is
coupled with the magnetization. As suggested by Brock \cite{brock}, 
understanding the dynamical evolution of such a class of systems
would be important in modelling price processes with complex systems.
The interesting property of Proposition \ref{expec} is that it does
not impose much restriction on the topology of spin
interactions under consideration. In fact, Proposition \ref{expec}
holds for very complex topologies of (local) interactions: for usual
lattice topologies or even more complex structured graphs of social
influences. The only essential assumption of Proposition
\ref{expec} is that $s_i(0)$, $i \in \Lambda$, have \emph{network
homogeneous mean}, and that the noise density of the noise
distribution $\Phi$ is uniform and symmetric around zero.

The reason for this independence of the network structure
is the proportionality of the social interaction of an agent to
the {\it relative} excess demand observed in her peer group.
Therefore the actual size of the peer group (as reflected in the
network structure) is not relevant for the decision of an agent.
In order to avoid confusion we stress, however, that the use of
the expectation value in the proposition does not automatically
imply that the typical behaviour of a single realization of
the process is described independently of the
network structure. This is guaranteed only after applying an
ergodicity argument for a given network. Then
equation (\ref{difeq}) describes ultimately the dynamics of the
external field (the dynamics of price) when the number of agents
goes to infinity.   In the next section we shall examine more closely
the behaviour of relative excess demand for
a single realization and also supply numerical evidence that, in
applications, our results remain valid with great accuracy at times 
beyond $t^\ast$.

%
%

\section{Collective behaviour}\label{application}

\subsection{Preliminaries}\label{how}

In the following two
Subsections~\ref{regone} and \ref{regtwo} we deduce
some corollaries of Proposition~\ref{expec} that reveal interesting patterns for the 
dynamics of the relative excess demand
and asset price in our model. We consider two particular choices of the fundamental
value dynamics (recall that $P_f(t) = U(P(t),t)$ as defined in {\ref{fund})):
\begin{eqnarray}
(A)&&\hbox{The average fundamental value evaluation of the asset remains
fixed all the time:}\nonumber \\
&& P_f(t) \equiv P_f(0) \hbox{ is a constant in time, and }
P(0)-P_f(0)=\gamma\hbox{ for some }\gamma\in \mathbb{R};
\label{sluca}\\
(B)&&\hbox{The \emph{difference} between the price and the average 
fundamental value evaluation}\nonumber\\
&&\hbox{remains fixed:}\nonumber \\
&& P(t)-P_f(t)=\gamma,\hbox{ for some
}\gamma\in \mathbb{R}\hbox{ and for all }t\geq 0.\label{difcon}
\end{eqnarray}
Since we are interested in short-time behaviour we do not
consider an explicit time-dependence of the parameter $\gamma$.

Our attention to these cases is motivated by a somewhat simplified
view according to which the fundamental value of an asset is
determined by the ``health'' of the firm that issued the asset. In
this view, an abrupt change of the fundamental value
may occur upon the arrival of news regarding the firm. It is then
plausible to admit that the average evaluation of the
fundamental value of the asset does not change in the time interval
between two consecutive arrivals. Such a real market situation is thus
appropriately modelled by case (A).
Another situation that might be realistic from our point of view, is
when the lack of information about the firm's health makes investors
infer the fundamental value of the asset from its market price. For
this situation, we consider that the individual asset evaluations 
will ``follow'' the asset price with a fixed,
but maybe non-zero difference $\gamma$. For positive [negative]
$\gamma$ this parameter represents a systematic mean overrating [underrating] 
of the market price from the point of view of the agents,
i.e., on average the
agents may consider the market to overrate [underrate] what each of them 
believes to be the actual fundamental value of the asset. 
This situation corresponds to the case (B).

Obviously, neither of the two situations described above can last for a
long time in a real market.  However, this fact does not constrain the
applicability of our results because the real market situation 
in question has
as its model counterpart the short-time behaviour to which our
proposition applies. (Mathematically speaking we mean by 
short-time behaviour the behaviour on a finite time interval
starting at $0$). 

Recall that Proposition 1 is concerned with the behaviour of the
expected relative excess demand $\mo[s_i(t)]$.  For notational brevity
we shall henceforth denote this quantity as $m(t)$ (our choice
of letter is inspired by the analogy with magnetism in a spin
system).  For a given choice of fundamental value dynamics,
Proposition 1 allows us to equate $m(t)$ on the interval $[0,t^*]$ with
the solution $f(t)$ of an ODE.  For the choices (A) and (B) above this
leads to the explicit results presented below as Corollaries 1 and 3 respectively.

Thus far our analysis has been mathematically rigourous.  However, in
order to make statements about the relative excess demand in a \emph{single
realization} we need to make an additional assumption.  The assumption is that for 
each $t\in [0, t^\ast]$ the random variable
\begin{equation}\label{close}
\bar{m}(t) = \frac{1}{|\Lambda|}\sum_{i\in \Lambda}s_i(t)
\end{equation}
is close to a constant that is equal to the common mathematical
expectation of each variable $s_i(t)$. This property would indeed
follow if the sequence $\{s_i(t), i\in \mathbb{N}\}$ obeyed the Law
of Large Numbers, or if it were an ergodic sequence. However, to 
justify the assumption in this way, we need to consider our model on an
infinite graph. Moreover, to use the ergodic theory, we would need
to impose that the peer groups are finite, of a finite range and of
a regular structure; this would allow us to construct a sequence of
imbedded graphs $\cdots \subseteq \mathcal{G}_n \subseteq
\mathcal{G}_{n+1} \subseteq \cdots $, which, in turn, would allow us
to formulate the ergodic property assumption in terms of the
convergence of $|\mathcal{G}_n|^{-1} \sum_{i\in
\mathcal{G}_n}s_i(t)$ to the common expectation of the $s_i(t)$'s.
The reader can find the details of this approach in, for example, \cite{horst}.
We did not pursue it here
because we found the mathematical formalism to be unnecessarily
cumbersome for our purposes.  In our work, the additional assumption
is instead ``verified'' 
by computer simulations (see Section~\ref{computer}).

The argument that $\bar{m}(t) \approx m(t)$ (with exact
equality in the limit of infinite system size) allows us to write down
an expression for the asset price $P(t)$ defined in~(\ref{evp}).
Specifically, under the foregoing assumption, we have
\begin{equation}\label{poso} 
P(t) = P(0)+\lambda\int_0^t m(u)du\hbox{ for }
t\in [0,t^\ast].
\end{equation}
where, by Proposition 1,
\begin{equation}\label{cone}
m(t)= f(t) \hbox{ for each }t\in[0,t^\ast] \hbox{ provided }f(0)=m(0).
\end{equation}
Moreover, $t^\ast$ may be calculated or estimated from the
relation
\begin{equation}\label{estast}
t^\ast=\min\left\{ t\geq 0\, :\, \left|P(0)+\lambda\int_0^t
m(u)du-P_f(t)\right|\leq \theta-\kappa\right\},
\end{equation}
which is equivalent to (\ref{tast})).

Corollaries 2 and 4 below present conclusions about the behaviour of
the asset price obtained using (\ref{poso}) for the cases (A) and (B) respectively.

\subsection{Case A: Fixed average fundamental value evaluation}
\label{regone}

Applying the ideas described in
Section~\ref{how}, we get the following two corollaries. The
formulation employs the following shorthand notations: 
\bel{short}
a:=\frac{1-\kappa/\theta}{2},\quad
b:=\sqrt{\frac{\lambda}{\theta}-\left(\frac{1-\kappa/\theta}{2}\right)^2},\quad
c:=\sqrt{\left(\frac{1-\kappa/\theta}{2}\right)^2-\frac{\lambda}{\theta}}.
\ee

\begin{description}
\item{\bf Corollary~1.} 
{\em If the average fundamental value evaluation is constant in time,
i.e., if the condition (\ref{sluca}) holds, the time pattern of
the expected relative excess demand, $m(\cdot)$, depends on the sign of the
determinant expression $1-\kappa/\theta-2\sqrt{\lambda/\theta}$ and
is for $t\in [0,t^\ast]$ as follows:
\begin{description}

\item{(a) If $1-\kappa/\theta-2\sqrt{\lambda/\theta}>0$} 
then
\begin{equation}\label{sola}
m(t)=\frac{1}{2c}\left[e^{-(a+c)t}\left(
m(0)(a+c)+\frac{\gamma}{\theta}\right)
-e^{-(a-c)t}\left(m(0)(a-c)+\frac{\gamma}{\theta}\right)\right];
\end{equation}

\item{(b) If $1-\kappa/\theta-2\sqrt{\lambda/\theta}=0$} then 
\begin{equation}\label{cara}
m(t)=\left[m(0)-\left(a m(0)+\frac{\gamma}{\theta}\right)t\right]e^{-a t};
\end{equation}

\item{(c) If $1-\kappa/\theta-2\sqrt{\lambda/\theta}<0$} then
\begin{equation}\label{solformb}
m(t)=e^{-at}\left\{m(0)\cos(bt)+\frac{1}{b}\left[-\frac{\gamma}{\theta}-am(0)\right]\sin(bt)\right\}.
\end{equation}
\end{description}
}
\end{description}

This Corollary arises from (\ref{cone}), Proposition~1 and (\ref{sluca}) by
noting that (\ref{difeq}) may be rewritten as a second-order
ODE
\bel{odegen}
f^{\prime\prime}(t)+(1-\kappa/\theta)f^\prime(t)+\left(\lambda/\theta\right)f(t)=0.
\ee 
This is the evolution equation for the damped harmonic oscillator.
The well-known general form of the solution of (\ref{odegen}) is
\begin{equation}\label{sol-again}
f(t)=\left\{\begin{array}{rl}
\left\{\left(1+t(1-\kappa/\theta)/2\right)f(0)+f^\prime(0)t\right\}e^{-t(1-\kappa/\theta)/2},&
\hbox{if }1-\kappa/\theta=2\sqrt{\lambda/\theta},\cr
\frac{-\left\{f^\prime(0)+\delta^- f(0)\right\}\mbox{e}^{-\delta^+t}
+ \left\{f^\prime(0)+\delta^+
f(0)\right\}\mbox{e}^{-\delta^-t}}{\delta^+ - \delta^-}, &\hbox{
otherwise }\end{array}\right.
\end{equation}
where (below $i=\sqrt{-1}$) \bel{del-again} \delta^\pm = \left\{
\ba{ll} (1-\kappa/\theta)/2\pm
\sqrt{\left\{(1-\kappa/\theta)/2\right\}^2-\lambda/\theta}& \mbox{
if }
1-\kappa/\theta> 2\sqrt{\lambda/\theta}, \\
(1-\kappa/\theta)/2\pm i
\sqrt{\lambda/\theta-\left\{(1-\kappa/\theta)/2\right\}^2}& \mbox{
if } 1-\kappa/\theta<2\sqrt{\lambda/\theta}. \ea\right. 
\ee 
With the initial conditions
\begin{equation}\label{icode}
f(0)=m(0)\hbox{ and }f^\prime(0)=-m(0)\left(1-\kappa/\theta\right)-
\gamma/\theta
\end{equation}
the corollary follows.

Corollary~1 provides us with precise information about the short-time
behaviour of the expected relative excess demand in our model.  Under
the additional assumption that $\bar{m}(t) \approx m(t)$, this also
yields information about the asset price (\ref{evp}). By close detailed inspection of the
solutions for the three cases we may rephrase Corollary~1 
as applied to the price as follows.

\begin{description}\item{\bf Corollary~2.}
{\em If the average fundamental value evaluation is constant in time, i.e., when the
condition (\ref{sluca}) holds, the pattern of the asset price,
$P(\cdot)$ on the time interval $[0, t^\ast]$ depends on the sign
of the determinant expression
$1-\kappa/\theta-2\sqrt{\lambda/\theta}$. In particular,
\begin{description}\item{(a) If $1-\kappa/\theta-2\sqrt{\lambda/\theta}>0$} then the price is either
a monotonic function, or exhibits a single extremum;
\item{(b) If $1-\kappa/\theta-2\sqrt{\lambda/\theta}=0$} then the price is a monotonic function;
\item{(c) If $1-\kappa/\theta-2\sqrt{\lambda/\theta}<0$} then the price is a $\sin$-like function
damped by an exponential function;  there is a large set of
parameter values for which more than one oscillation of this function
fit into the time interval $[0, t^\ast]$.
\end{description}
}
\end{description}
This result for the number of extrema in each of the three cases 
is proved in the appendix. It is non-trivial because of the limitation
to the finite time-interval $[0,t^\ast]$.

To demonstrate the significance of the corollaries we first note that the
parameter $\gamma$ has no influence on the qualitative properties discussed
in Corollary~2.
In the following we analyse how each of the parameters $\kappa,
\theta$ and $\lambda$ affects the fluctuation pattern of the asset
price in the non-equilibrium time range $[0,t^\ast]$. This analysis 
provides clear results -- albeit not trivial --
because of a particular feature of the determinant expression: 
changing the value of one of
the parameters  while keeping all other parameters fixed, the expression
changes its sign exactly once. The borderline case 
$1-\kappa/\theta-2\sqrt{\lambda/\theta}=0$ is
excluded from our consideration because obviously one cannot expect a
real market to fit perfectly our model and have parameter values 
fitting exactly this equation.

\medskip\noindent{\bf Social susceptibility:} 
Here we consider the dependence on the parameter $\kappa$ which
reflects the strength of the social susceptibility.  Note that, by
definition, $\kappa$ is non-negative and that, from~(\ref{tast}), the existence of positive $t^*$ implies $\kappa <
\theta$.  Hence the value range for $\kappa$ to be considered here is $[0, \theta)$. The consideration
splits into two cases. The first case is
$1<2\sqrt{\lambda/\theta}$. In this case, the inequality
$1-\kappa/\theta-2\sqrt{\lambda/\theta}<0$ holds for all values of
$\kappa$, and, according to Corollary~2, the price fluctuates
indefinitely. In the opposite case, $1>2\sqrt{\lambda/\theta}$, we can
find a critical value $\kappa_c(\theta, \lambda)\in [0, \theta)$
such that $1-\kappa/\theta -2\sqrt{\lambda/\theta}>0$ where
$\kappa<\kappa_c$ while $1-\kappa/\theta-2\sqrt{\lambda/\theta}<0$
leads to $\kappa>\kappa_c$. By the corollary, this implies that the
price may bounce at most once when the social influence is weak,
while it oscillates indefinitely when the social influence is strong.

This conclusion is not a totally novel discovery since it may be
deduced by a simple argument which is well known from interacting
agent models for asset markets (see \cite{lux}): the strong social
influence causes the herding behaviour, which expresses itself in
that an agent buys (resp., sells) when the majority buy [resp.,
sell], so that the price grows (resp., falls) beyond the fundamental
value. When the asset price is much beyond the fundamental value
the fundamentalist trading component makes the agents change their
positions, which reverses the price trend. Interestingly, the exact
analysis of our model suggests that a market asset price may be forced 
to oscillate by strengthening solely the social influence.

The main novelty of our conclusion is that it indicates that
{\em the price may bounce when the interaction between agents is low
or even when it is totally absent}.  Another novel aspect is that {\em
under weak interactions the price can bounce no more than  once,
whereas under strong interactions it may oscillate}. This
alternative could hardly be explained by a heuristic argument on the
basis of the herding effect.

\medskip\noindent{\bf Heterogeneity of agents:} Here we consider the 
influence on the price pattern
of the value of  $\theta$, the parameter that expresses the
heterogeneity of the agents.  Due to the constraint (\ref{tast}),
the value range to be considered is $(\kappa, +\infty)$. Clearly,
for any $\kappa$ and $\lambda$, there is a finite critical value
$\theta_c(\kappa, \lambda)$ within this range such that
$1-\kappa/\theta-2\sqrt{\lambda/\theta}$ is negative if
$\theta<\theta_c$, and is positive if $\theta>\theta_c$. By
Corollary~2, this means that {\em a low heterogeneity of
agents may cause the price to oscillate many times, while under a
high heterogeneity, the price can bounce only once}. This effect
agrees with a traditional explanation of how the heterogeneity
influences the ups and downs of price. The essence of that
explanation is the following. If the heterogeneity is low almost all 
agents have almost the same
individual opinion with respect to the fundamental value. Suppose now
that the asset price is not very high. Then almost all agents wish
to buy if they obey their fundamentalist trading strategy. The
social interaction makes these ``wishes'' even stronger. Thus a
majority will be buying which makes the price grow.
It will be growing until it becomes higher than the maximal
evaluation of the fundamental value of the bulk of the population.
At this moment, the situation is inverted and the predominant
majority will be selling. Thus the fluctuations of the demand and
price arise. This line of reasoning has been employed by various authors 
to justify a phenomenon that is phenomenologically similar to the effect 
described by our model, but concerns the long-time fluctuations of the price.
However, this qualitative explanation
could not distinguish whether the up-and-down would occur once or
several times.

\medskip\noindent{\bf Market feedback:} Here we discuss the influence 
of the parameter $\lambda$ on the asset price function, again paying
particular attention to whether this function can bounce once or more.
To the best of our knowledge there is no analytically tractable
HIAM for which this aspect of a market has been studied thoroughly.

We could argue as in the previous discussions and deduce that
{\em for any $\kappa$ and $\theta$, there is a critical value
$\lambda_c(\kappa, \theta)$ such that when $\lambda<\lambda_c$ the
price can bounce at most once, while when $\lambda>\lambda_c$ the
price can oscillate}. However, we would like to concentrate our
arguments on another conclusion that we found to be deeper, more
important and striking: {\em it is the presence of the linear
feedback of the relative excess demand on the price increment that
enables the price to behave in time in two different ways; either it
is a damped $\sin$-like function, or it is a function that may have 
at most one extremum}. We note that we do not claim that only linear
feedback is able to cause such a dichotomy. A complex time behaviour of
the price function is a general consequence of the circular relation
chain
``price$\rightarrow$decisions$\rightarrow$demand$\rightarrow$price''.
Our conclusion concerns the ``demand$\rightarrow$price'' link,
saying that in this chain it is the main reason for the
dichotonomy of the price time behaviour. One
more factor in its favour will be given in the next subsection.

\subsection{Case B: Systematic market overrating}
\label{regtwo}

We now consider the second scenario for $P_f(t)$ where, on average,
the traders assume that the fundamental value of the asset and the market
price follow each other, but in their average opinion there may be
some constant offset $\gamma$ between the two quantities. We
refer to overrating (corresponding to positive $\gamma$) throughout
even though $\gamma$ may take negative values, with the obvious
interpretation that negative overrating means underrating.
Since under real circumstances a constant systematic overrating 
cannot be expected to be very ``large'' in any sense, we arrive at

\begin{description}\item{\bf Corollary~3.} {\em
In the case of systematic market overrating,
that is, $P(t) - P_f(t) = \gamma$ for a constant $\gamma$ for
all $t\geq 0$, the condition
\begin{equation}\label{ogra}
\left|\gamma\right|\leq \theta-\kappa
\end{equation}
ensures that the expected relative excess demand on the time
interval $[0,\infty)$ is given by
\bel{4-1} m(t) = \left(m(0)
+
\frac{\gamma}{\theta-\kappa}\right)\mbox{e}^{-t\left(1-\frac{\kappa}{\theta}\right)}
- \frac{\gamma}{\theta-\kappa}, 
\ee 
where $\theta > \kappa \geq 0$.}
\end{description}

From (\ref{evp}), combined with the assumption $\bar{m}(t) \approx m(t)$, we then obtain
\begin{equation}\label{ogra2}
P(t) = P(0) + \frac{\lambda\theta}{\theta-\kappa} \left(m(0) - m^\ast\right)
\left(1-\mbox{e}^{-t\left(1-\frac{\kappa}{\theta}\right)}\right) + \lambda m^\ast t
\end{equation}
where the limit relative excess demand
\begin{equation}\label{minf}
m^\ast :=  m(\infty) = - \frac{\gamma}{\theta-\kappa}
\end{equation}
determines the limit growth rate $\lambda m^\ast$ of the asset price.
Hence rephrasing Corollary 3 in terms of the asset price
and analysing the number of extrema, we find

\begin{description}\item{\bf Corollary~4.} {\em If a systematic
market overrating prevails, that is, if $P(t) - P_f(t) =
\gamma$ for a constant $\gamma$ for all time, then the additional
condition (\ref{ogra}) ensures that $t^\ast=\infty$ and the
asset price of the model behaves on $[0, \infty)$ as follows:
\begin{equation}\label{pcase2}\begin{array}{rcl}
\hbox{it is a monotonic function in time}& &\hbox{if }\sign(m(0))\not= \sign(\gamma),\\
\hbox{and it bounces once}& &\hbox{if }\sign(m(0))=\sign(\gamma).
\end{array}\end{equation}
}
\end{description}

Before discussing the significance of the model parameters in this setting we remark
that the unbounded growth (or decay) of the price is correct for the model, but
not claimed to exist in a real market since the model assumption $P_f(t) = P(t)-\gamma$
cannot remain true for a long period of time. 

\medskip\noindent{\bf Market Feedback:} 
To illustrate the significance of this second set of corollaries we first note that the
feedback parameter $\lambda$ has no influence on the qualitative properties discussed
in Corollary~4. This demonstrates that if strategies 
{\it on average} assume fundamental value and market price to match (up to some 
constant bias), then the feedback between price and relative excess 
demand becomes irrelevant for the behaviour of the market.
As a consequence, the asset price cannot exhibit more than one bounce.
This observation
strengthens our previous conclusion that this feedback is the main factor 
that determines whether the asset price can oscillate or not.

This might appear surprising at first sight, but it can be easily derived 
from the mathematical treatment of the model.
If the average evaluation of the fundamental value
follows the market price, the decision rule (\ref{as})
of the agents takes the following form:
\begin{equation}\label{buysellR}
\hbox{ if \ \ }\frac{\kappa}{|J_i|}\sum_{j\in J_i}s_j(t)-
\bigl[\gamma+x_\ell^{(i)}\bigr]\left\{\begin{array}{c}\geq 0\\
< 0\end{array}\right.\hbox{ then \ \
}s_i(t)=\left\{\begin{array}{l}+1\\ -1.\end{array}\right.
\end{equation}
Observe that neither the asset price nor the
average evaluation of the fundamental value of the asset appear in
(\ref{buysellR}). The connection ``price$\rightarrow$decisions'' 
from the generic circular relation
chain ``price$\rightarrow$decisions$\rightarrow$demand$\rightarrow$price''
is manifestly broken. Mathematically this causes the absence of the asset 
price in the ODE (\ref{difeq}) (for the link between the rule (\ref{as}) and
(\ref{difeq}), see the proof of Proposition~\ref{expec} in the
appendix). Then this ODE takes the following form:
\bel{dfgama}
f^\prime(t)=-\left(1-\frac{\kappa}{\theta}\right)f(t)-\frac{\gamma}{\theta}.
\ee
This is an ordinary differential equation of
first order and it is easy
to check that in the range of the allowed parameter values,
its solution has at most one root, i.e., at most one $t_0$ for
which $f(t_0)=0$. Thus, since $f(\cdot)$ is proportional to the derivative 
of the asset price function, the asset price cannot have more than one
extremal point.

For the remaining model parameters we note that only the combinations
$\kappa/\theta$ and $\gamma/\theta$ occur in (\ref{dfgama}) and thus
we need not explicitly consider the dependence on $\theta$.

\medskip\noindent{\bf Social susceptibility:} 
We remind the reader that throughout our treatment the condition
$\kappa<\theta$ for mild herding behaviour is assumed.
Corollary~3 tells us that 
the expected relative excess demand is a monotonic function in
time, and that
its limiting equilibrium value is (\ref{minf}).
This means that in the equilibrium limit of our model, the proportion of 
those who buy exceeds the proportion of
those who sell by $\frac{-\gamma}{\theta-\kappa}$. In our non-equilibrium
setting, however,
it is more interesting to ask how fast $m(t)$ converges to its limit.
The answer is in Corollary 3: the larger the social susceptibility
relative to the heterogeneity $\theta$, the slower is this convergence.
This gives some insight into how the heterogeneity and the social influence 
compete in a market.

\medskip\noindent{\bf Market overrating:}
The sign of $\gamma$ determines whether in the long run the price will be 
growing or falling; asymptotically $P(t) \sim \lambda m^\ast t$. Since
therefore $P(t)$ can become negative, it may be useful to remind the
reader that $P$ really means the logarithm of the price, and that the
asymptotic linear behaviour corresponds to exponential growth or loss
of the value of the asset with rate $1/(\lambda m^\ast)$.
Note that this does not depend on the
initial relative excess demand $m(0)$. However, the initial distribution
does play a role for the short-time non-equilibrium behaviour, since
to first order in time $P(t) = P(0) + \lambda m(0) t$.
The dichotomy
(\ref{pcase2}) of Corollary~4 agrees with the general understanding of how the
fundamentalist trading effect might affect the dynamics of the price.
Indeed, $\sign(\gamma)$ determines whether the majority ``mood'' is
to buy or to sell if the social interaction is absent.  The
social interaction by itself makes an agent agree with the
majority's mood. In our model, the initial majority mood is given by
$\sign(m(0))$. Thus, if $\sign(m(0))\neq \sign(\gamma)$ then our
heuristics suggests that the majority mood must be same all the
time. In fact, if $\sign(\gamma)=-1$, then $P(t)<P_f(t)$, i.e., the
price is less than the average evaluation of the fundamental value,
accordingly, the majority is inclined to buy, that is, $\sign(m(t))
\equiv 1$. On the contrary, if $\sign(\gamma)=1$, then the
price is higher than the average evaluation of the fundamental
value, so that the majority is inclined to sell, that is,
$\sign(m(t)) \equiv -1$. This suggestion is confirmed by the
corollary.

The opposite case, where $\sign(m(0)) = \sign(\gamma)$,
is more tricky in the sense that it is difficult to predict
from a purely heuristical argument
whether the initial mood will pertain for ever or whether it will be reversed.
Our model suggests that fundamentalist thinking will eventually
win over any initial mood, even if the herding effect (which
slows down the relaxation to the asymptotic behaviour) tends
to preserve the initial mood for some time. The time of mood reversal
(i.e., when $P(t)$ reaches again $P(0)$)
grows with increasing social susceptibility.


\subsection{Computer simulation tests}\label{computer}

To complement the preceding theoretical analysis, we here present the
results of Monte Carlo simulations using random sequential update
as discretization of the continuous-time process.  We
first discuss results corresponding to the case with fixed fundamental
value (model (A)) before treating the systematic market rating case
(model (B)).  In both cases we choose some specimen parameter values
and compare the observed behaviour of the relative excess demand to that
described by the solution of the ODE (\ref{difeq}).  As demonstrated
above, this solution is exact only in the time interval $[0,t^*]$ given
by~(\ref{estast}).  However, for a large ensemble of strongly-connected traders
we expect it also to be a good approximation outside this regime. To
test this assumption we deliberately choose the worst-case scenario
with parameter values so that $t^*=0$. We consider fully-connected networks
and undirected random networks with a constant small connectivity 
$\nu$ which is the number of peers of each agent.

For model (A) we concentrate on the case of strong feedback and strong
social susceptibility at
the point, $\lambda/\theta > a^2 =0$, for which the
treatment of section~\ref{regone} predicts \emph{undamped} oscillations of the
excess demand.  We note that periodic oscillations were previously
seen in simulations of a different model combining price adjustment
and interactions between heterogeneous agents~\cite{Weisbuch04}.

Figure~\ref{f:Bfullysing}
shows results for a single history on a large fully-connected
network of agents.  We see immediately that even a single history is well
described by the ODE~(\ref{difeq}) with, for large system sizes, only a slight
fluctuation in the amplitude of the oscillations.  The average over
histories shown in Fig.~\ref{f:Bfullyave} is essentially
indistinguishable from the theoretical prediction of~\eqref{solformb}
with $a=0$. 

\begin{figure}
\begin{center}
\mbox{\subfigure[Single
realization]{\includegraphics*[width=0.49\textwidth]{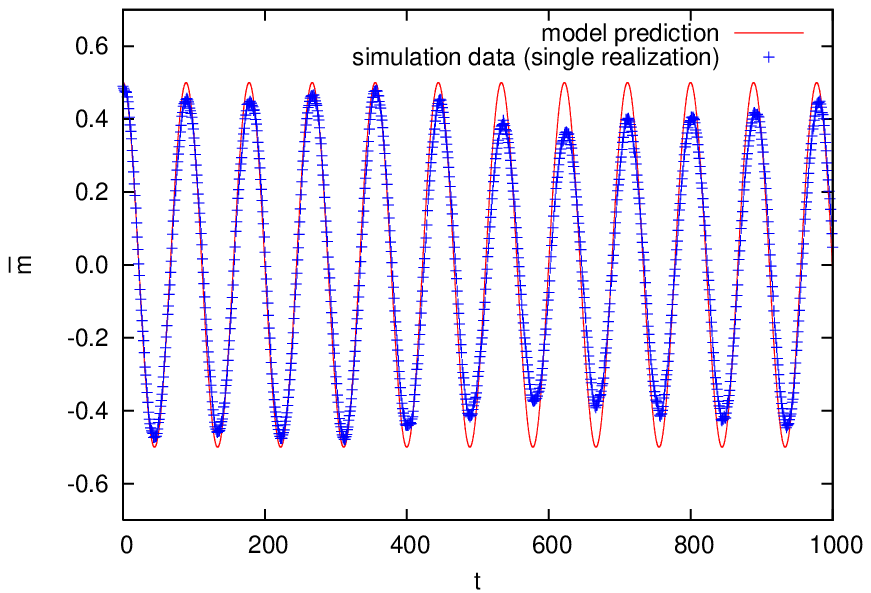}
  \label{f:Bfullysing}}\quad
  \subfigure[Average over 100
  histories]{\includegraphics*[width=0.49\textwidth]{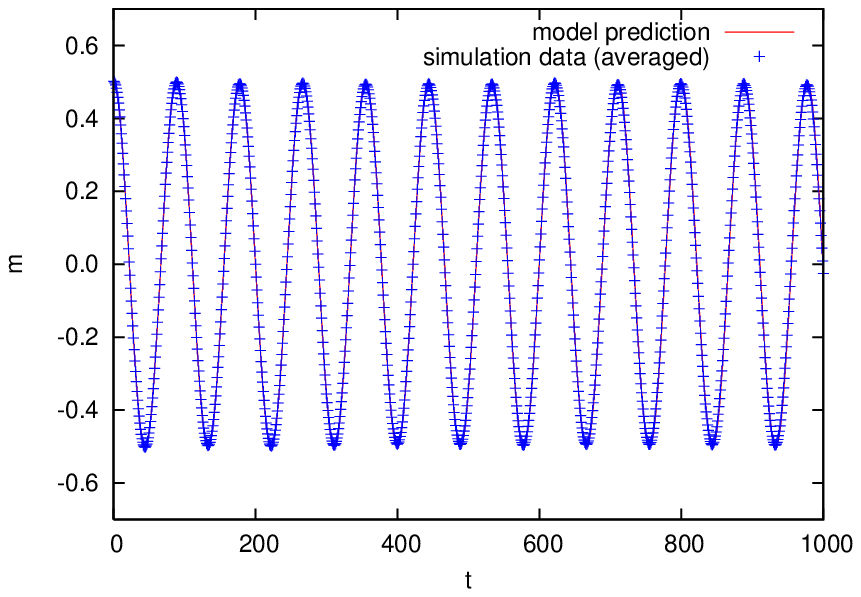}
  \label{f:Bfullyave}}}
\caption{Relative excess demand as a function of time for model (A) with 10000
  agents on a fully-connected network. In (a) we show the random
  number $\bar{m}(t)$ (\ref{close}) for
  a single realization, whereas in (b) we show an average of this
  quantity over 100 histories which gives an estimate of $m(t)$.
  Crosses are simulation data; solid red line is the model prediction
  of~\eqref{solformb}.  System parameters: $\lambda=0.5$, $\theta=1$,
  $\kappa=1$, $P(0)=P_f(0)=1$.
}
\label{f:Bfully}
\end{center}
\end{figure}

The situation for networks with lower connectivity is more subtle, as
demonstrated in Fig.~\ref{f:Bsmall} 
for an undirected random network
with connectivity small relative to the system size.
Oscillations are still clearly seen but the frequency is slightly
lower than predicted and there is an initial decay before fluctuation
about some lower amplitude.  (For the averaged case the amplitude
decreases still further due to fluctuations in the frequency for each
history.)  We believe this effect is a result of the build-up of
correlations---a hypothesis which is supported by observations that
the discrepancy from prediction is larger for a nearest-neighbour
network (with the same connectivity) but smaller for a directed
network. Real trader networks would be expected to have small-world or
scale-free properties but
to show qualitatively the same behaviour.  A more detailed account
of the dynamics of our model on such networks will be published elsewhere.

\begin{figure}
\begin{center}
\mbox{\subfigure[Single realization]{\includegraphics*[width=0.49\textwidth]{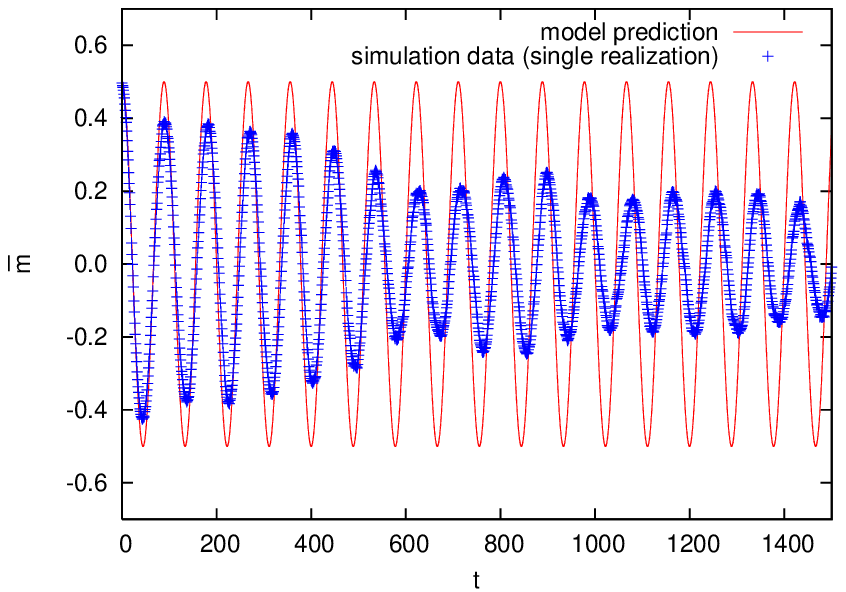}
  \label{f:Bsmallsin}}\quad
  \subfigure[Average over 100 histories]{\includegraphics*[width=0.49\textwidth]{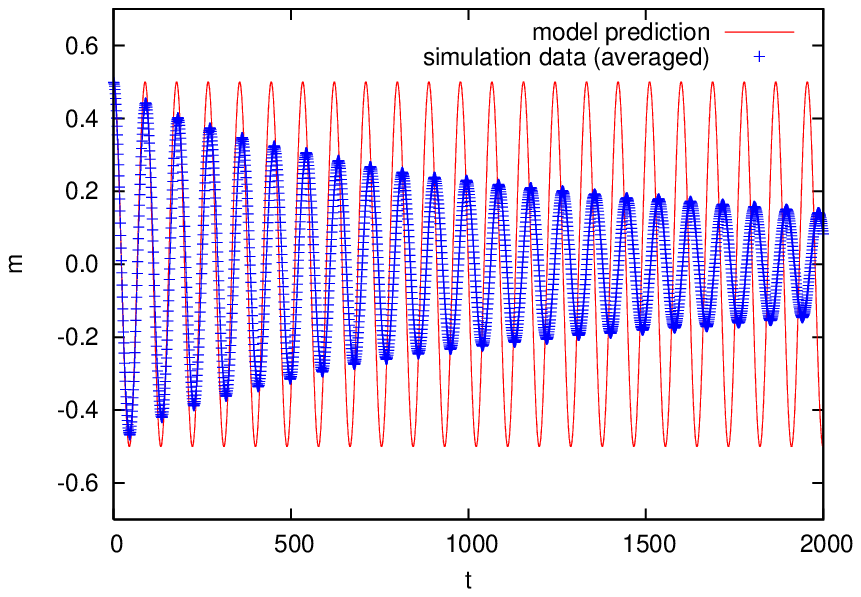}
  \label{f:Bsmallav}}}
\caption{Same as Fig.~\ref{f:Bfully} but for a random undirected
network ($L=10000$) with fixed connectivity $\nu=6$. In other words, each agent has exactly six 
peers drawn randomly from the community with the constraint that if agent $j$ is a peer 
of $i$ then agent $i$ is a peer of $j$.}
\label{f:Bsmall}
\end{center}
\end{figure}

We have also simulated the overdamped case (i.e., $\lambda/\theta >
a^2  >0$).  Here again, results for a fully-connected network show
excellent quantitative agreement with the theoretical
prediction~\eqref{solformb} but there is a slight discrepancy for
less-connected networks.  However, in this case there appears to be
less dependence on the network structure for given connectivity.

Let us now turn to the systematic market rating case of model (B).  As
argued at the start of section~\ref{regtwo}, if the condition
$\left|\gamma\right|\leq \theta-\kappa$ is obeyed then $t^*=\infty$.
On the other hand, if this condition is not fulfilled, it is
straightforward to show that $t^*=0$. In this
case the following heuristic picture emerges for a strongly connected network where the average
number of trusted friends of each trader is of the order of the
network size
$\Lambda$. Since we expect $\bar{m}(t) \approx m(t)$, up to
fluctuations of order $1/\sqrt{\Lambda}$, we argue that the solution
(\ref{4-1}) should be approximately valid until some time $t^{**}$ determined by the condition $|\kappa
m(t^{**}) + \gamma| = 1$. From then on all traders opt to buy
(if $\kappa m(t^{\ast\ast})+\gamma = -1$) or sell
(if $\kappa m(t^{\ast\ast})+\gamma = 1$). Consequently the system continues to
evolve according to
\bel{freeze}
\frac{d}{dt} m(t) = - m(t) \pm 1
\ee 
with unit relaxation time which is just the intrinsic decision
timescale of the continuous-time dynamics. In other words, after the
crossover time $t^{\ast\ast}$ the system begins quickly to freeze in
the non-fluctuating equilibrium state, i.e., the uniquely-given frozen
state with $m^\ast = \sign{(-\gamma)}$.

Figure~\ref{f:A}
\begin{figure}
\begin{center}
\mbox{\subfigure[Fully-connected
  network]{\includegraphics*[width=0.49\textwidth]{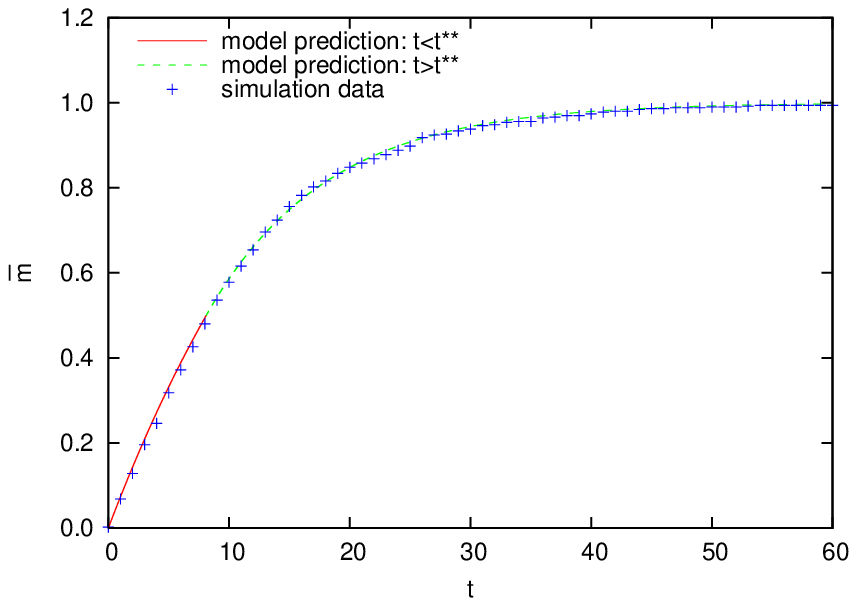}
  \label{f:Afull}}\quad \subfigure[Low-connectivity random
  network]{\includegraphics*[width=0.49\textwidth]{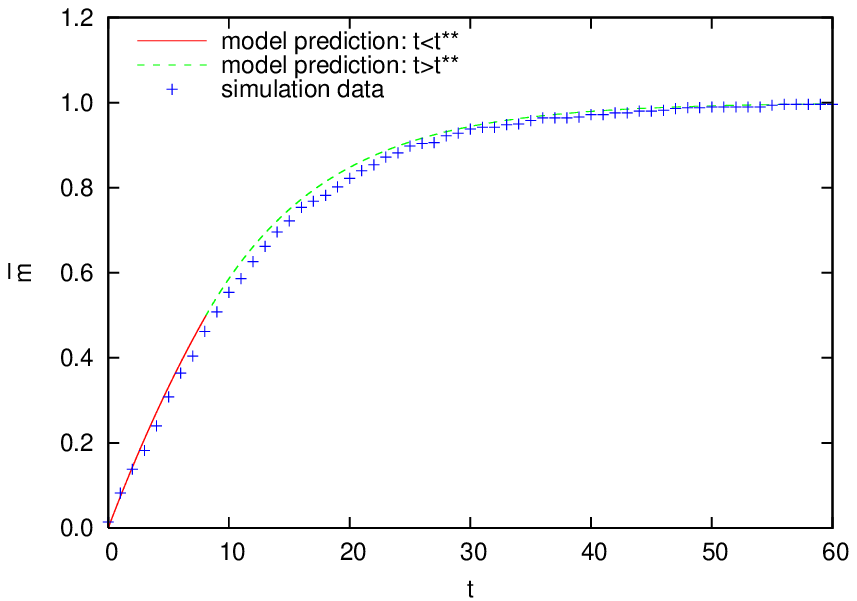}
  \label{f:A6}}}
\caption{Relative excess demand as a function of time for model (B) with 1000
agents on (a) a fully-connected network and (b) a random undirected
network with fixed connectivity $\nu=6$.  In both cases the graph
shows $\bar{m}(t)$ for a single realization against $t$. Crosses are
simulation data; lines are the model predictions of Eqs.~\eqref{4-1}
and~\eqref{freeze} for $t<t^{**}$ (solid red) and $t>t^{**}$ (dashed green)
respectively.  System parameters: $\lambda = 1$, $\theta = 1$, $\kappa = 0.5$,
$\gamma = -0.75$, $P(0)=0$. }
\label{f:A}
\end{center}
\end{figure}
shows results of simulations for model (B) with $-\gamma > 1-\kappa > 0$
(small social heterogeneity, large market under rating) on two different
networks.  As predicted we find that the
relative excess demand evolves towards an equilibrium value of
$\sign(-\gamma)$.  In fact, the evolution of $\bar{m}(t)$ for a single
history of the process is very well described by the predictions of
Eqs.~\eqref{4-1} and~\eqref{freeze} for the expectation $m(t)$.
The agreement is excellent in the fully-connected case but still good
when the connectivity is low compared to the system size.  As
expected, fluctuations were found to be larger for smaller system
sizes, these fluctuations can be smoothed out by averaging over a
number of histories.

\subsection{Peer network}
\label{network}

A perhaps surprising result is the equality of the expectation
of decision $\mo{s_i(t)}$ for all agents $i$ for any network
provided that the initial distribution was network mean homogeneous.
We point out that for a fully connected network this is less
surprising. To see this, imagine a discrete-time version of the
model where all agents are updated simultaneously in one time-step
following the decision rule of our model.
Then, in order to initiate our model, that is, in order
to be able to construct the opinion distribution over
the population at time $t=1$, it is sufficient to give
$m(0)$ -- the  relative populational excess demand
at time $0$. Then, in accordance with our model definition,
the measure on $\{s_i(1), i\in \Lambda\}$ will be
network mean homogeneous. Consequently, the measure that
rules $\{s_i(t), i\in \Lambda\}$ for any $t>1$ will also be
network mean homogeneous.

In other words, in the fully connected mean-field case of our
model, the distribution of the agent opinions is network mean
homogeneous, when the model is initiated as described above.
The opposite case for initiating the model is to give the
distribution of $\{s_i(0), i\in \Lambda\}$. This distribution
may have non-homogeneous mean.  But since each agent is
influenced by other agents solely through the relative
populational demand, then the measure on $\{s_i(1),i\in \Lambda\}$
will already have network homogeneous mean after one time step.
Therefore, for fully connected networks only network mean
homogeneous measures are of interest.

For a fully connected network we can also
explain the numerical observation that for a large
population a single realization of the process essentially coincides with
the analytically computed mean.
One may invoke the law of large numbers for the peer excess demand
to argue that
the social interaction becomes deterministic. Then the stochastic
dynamics is solely driven by the uncorrelated annealed randomness of
the individual evaluation of the fundamental value and, from
ergodicity arguments, one expects individual histories to
follow the analytical result for the expectation.

A greater surprise then is the observation that individual realizations of the process
capture the main features of the average over histories even for
random networks with {\it small} connectivity. For small connectivity the
first step (large numbers for the peer excess demand) of the previous reasoning fails 
and it remains unclear why also histories
on graphs with small and heterogeneous connectivity are reasonably
well described by the analytical results. We rationalize this observation
by closer inspection of the two-fold origin of external randomness in the model, viz.
(i) the quenched random graph structure representing the (highly heterogeneous)
peer groups of each agent, and (ii) the annealed randomness of the individual
evaluation of the fundamental value. We note that in each individual
decision the randomness coming from the graph structure
adds to the randomness of the individual evaluation of the fundamental
value an amount of at most the same order (since $\kappa < \theta$)
and hence does not qualitatively change the
strength of the total external noise that drives the stochasticity of the model. Making the natural
assumption that the uncorrelated annealed noise of the individual evaluation of the
fundamental value dominates over the quenched noise then explains why the
``mean-field'' model of a fully connected network approximates quite well a
weakly-connected quenched random network.

We stress that our observations should not be interpreted as suggesting
that network topology would generally not matter for anything of interest
in the dynamics of general HIAMs.
Many studies of the Ising model on a variety of networks have revealed
interesting network phenomena, for example, for recent work
on Barab\'asi-Albert networks see \cite{Sumour05,Lima06} and more generally
on phase transitions and critical behaviour in networks, see e.g.
\cite{Lux99,Aleksiejuk02,Leone02,Goltsev03}.
Very recent mean-field and numerical work on a HIAM which is constructed
in a spirit similar to ours (similar in general terms, but quite different in its details) 
has shown that its equilibrium finite-size effects depend
strongly on the topology of the underlying network \cite{x}.
It would be very interesting to see whether the presence of phase transitions
and strong finite-size effects in the equilibrium distribution is reflected
in the short-time dynamics of stochastic models with these equilibrium
properties.

\section{Some further issues}
\label{Conclusions}

In summary, we have presented a heterogeneous interacting-agent model for the
short-time non-equilibrium dynamics of an asset price. This model
is analytically tractable and thus provides detailed insight into
the interplay of individual fundamentalist heterogeneity and social
interaction leading to herding behaviour for a variety of scenarios
of the relation between market price and average evaluation of the
fundamental value of the asset. Quite independently of the underlying 
structure of the peer network, interesting collective behaviour
emerges, including oscillations, a single bounce, or monotonic evolution
of the price.  These features can be traced back to the relative strength
of the factors that determine the time evolution of the HIAM, as discussed in 
detail in Sec.~\ref{application}.

An open question with respect to the random graph structure
of the peer network is the robustness of the assumption of domination of
the collective stochastic dynamics by annealed noise that explains
why our results have a fair degree of independence of the topology of the peer groups.
There is indeed much further scope for simulation studies looking at different
aspects of our model on more complex networks. The present results for simple
topologies confirm the relevance of our analytical approach and
provide a starting point for future work probing systematically various
classes of network topologies for correlations,
inhomogeneous initial distributions, and finite-size effects in these
quantities. One may even attempt to study its stationary distribution for
gaining insight in the relation between stationary and early-time
dynamical properties.

An anonymous referee pointed out that 
while our manuscript was under review for publication, there appeared work 
of Horst and Rothe \cite{horst2} that seems to be closely related 
to ours. Indeed, our Proposition~1 and Corollaries~1 -- 4 appear to be very similar 
to conclusions of Horst and Rothe that may be summarized as follows: The trajectories of the price
in the model of Horst and Rothe from \cite{horst2} converge to a solution of a 
differential equation (Theorem~2.5 from \cite{horst2}) and, depending on choice of 
parameters of the model,  this solution looks like that of the damped harmonic oscillator, 
or oscillates almost chaotically, or ``bounces'' once and then  converges exponentially 
fast to a constant (the examples of Section~3 from \cite{horst2}). 

Let us point out the
differences between our work and that of Horst and Rothe. The first is that in our model, 
the agents can have different peer groups while in the model of Horst and Rothe ``... all 
other individuals influence one particular trader in the same way'' (the citation from Remark~2.2 
of \cite{horst2}). Second, the feedback of excess demand on the price increment is a free 
parameter in our model while it is a constant (equal to $1$) in \cite{horst2}. This may 
be not a principal difference since one can normalize all the free parameters by the 
feedback value, but such a normalization would make impossible the analysis 
that we make in Sections~4.2 and 4.3 concerning how this feedback affects the evolution 
pattern of asset price. Next, in Horst and Rothe's model, agents can have different 
strategies that depend on the states of process in the past. This structure 
is much richer than that of our model since where an agent decides whether her 
order is to sell or to buy on the basis of the state of the process at the moment of decision. 
Certainly, a dependence structure as general as that of Horst and Rothe requires the imposition 
of constraints so as to allow for analytical treatment of the model. Horst and Rothe impose that 
all the order fluxes as well as the moments at which agents change their strategies are determined 
by Poisson point processes. In particular, every agent ``carries with herself'' {\it two} independent 
Poisson point processes, one of which determines the instances of the sell orders and the other one 
the instances of the buy orders. The rates of the processes change in time, but the processes are 
independent by the construction. This is different from our model: if we re-write the decision rule 
(described in Section~\ref{model}) in a form similar to that of Horst and Rothe then the resulting 
Poisson point processes will be dependent through the dependence of their rates. One more difference 
lies in the definition of how the price dynamics depends on the excess demand. In Ref.~\cite{horst2}, 
the price raises [resp., diminishes] when a buy [resp., sell] order is put. In our model, 
if an agent puts an order after she has put an order of the same type, the excess demand does 
not change and thus, the price continues to grow or decay at the same rate as it has been growing or 
decaying. 

It is worth mentioning that Horst and Rothe consider the law of the deviation of the 
stochastic trajectories from their limit (given by the solution to a deterministic differential equation); 
this issue is not addressed by us here. In summary, our model and that of Horst and Rothe are 
different, the result of neither of the two works follow from the results of the other one, and the 
methods used in the proofs are different. The similarity of results then suggests that they reveal a quite
general phenomenon, but we do not possess tools and results that would allow us to discuss this
suggestion in mathematical detail.

\section*{Acknowledgments}

This work was initiated through and supported by an DAAD/CAPES exchange 
grant (GMS and FP). We also acknowledge support by FAPESP (FP) and CNPq 
and FAPESP (VB). We thank M. Hohnisch, T. Lux and D. Stauffer for useful
comments on a preliminary version of the manuscript.

\appendix

\section{Proofs}
\label{appendix}

In this appendix we provide the details of two key proofs.  
Specifically, in Section~\ref{expec-proof} we prove Proposition~\ref{expec} and in
Section~\ref{col2p} we prove Corollary~2.  With respect to our other 
assertions, note that the proof of
Corollary~1 is outlined in Section~\ref{regone} of the main text;  Corollary~3
follows by a similar argument and Corollary 4 is a direct consequence
of Corollary~3.

\subsection{Proof of Proposition~\ref{expec}}\label{expec-proof}

The statement of Proposition 1 can be found in section~\ref{result}.
Here we shall deduce this result from a general property of continuous-time
Markov processes. In our case, the Markov process is 
$\{(s(t), P(t), P_f(t)),\, t\geq 0\}$ and its state space is denoted by
$\{-1, +1\}^{\Lambda}\times \mathbb{R}\times\mathbb{R}$.  
The finite-dimensional counterpart of the Hille-Yosida theorem 
(see \cite{liggett}, Chapter~I) then states that for a continuous and bounded 
function $g: \{-1, +1\}^{\Lambda}\times \mathbb{R}\times\mathbb{R}\, 
\rightarrow \, \mathbb{R}$
\begin{equation}\label{hy}
\frac{d}{dt}\mo\big[g\big(s(t),P(t), P_f(t)\big)\big]=\mo\big[\big(\mathcal{L}g\big)
\big(s(t), P(t),P_f(t)\big)\big],\kern1em \forall t\geq 0, 
\end{equation}
where $\mathcal{L}$ is the infinitesimal generator of the process
whose application to functions of interest to us
may be constructed as follows.

Let us fix arbitrarily a node $i$ from $\Lambda$, and let $g$ be the function that attributes
the value of $s$ at $i$ to any triple $(s, P, P_f)$:
\begin{equation}\label{fg}
g(s, P, P_f)=s_i.
\end{equation}
We now choose arbitrarily $(s, P, P_f)$ and let it be the initial state of our process. Since the times
of the updates of the position of any agent $j$ form a Poisson point process with rate $1$, and since 
the Poisson point processes related to different agents are independent, then the following is true:
during a short time interval $[0, u]$, the probability that the model's agent $i$ decides once 
to update her position is $u+o(u)$, the probability that she does not decide to update her
position is $1-u+o(u)$, and the probability of all other events is $o(u)$. Let us consider the
event when a single update decision is taken during $[0,u]$. Let $\tilde{u}$ 
denote the decision time ($\tilde{u}\in(0, u]$). Then, in accordance with the 
construction of our model, the sign of
$
\frac{\kappa}{|J(i)|}\sum_{j\in J(i)}s_j(\tilde{u})\, -\, \big\{P(\tilde{u})-\big(P_f(\tilde{u})+
X_i(\tilde{u})\big)\big\}
$
determines the resulting position. Thus, since $\Phi$ is the distribution function of the
random variable $X_i(\tilde{u})$
\begin{equation}\label{hiuv}\begin{array}{rcl}
s_i(u)=+1&\hbox{ with prob. }& 1-\Phi\left(-\left[\frac{\kappa}{|J(i)|}\sum_{j\in J(i)}s_j(\tilde{u})\, -\, 
\big\{P(\tilde{u})-P_f(\tilde{u})\big\}\right]\right),\\
s_i(u)=-1&\hbox{ with prob. }&
\Phi\left(-\left[\frac{\kappa}{|J(i)|}\sum_{j\in J(i)}s_j(\tilde{u})\, -\, 
\big\{P(\tilde{u})-P_f(\tilde{u})\big\}\right]\right).\end{array}
\end{equation}
Unfortunately,
we have no control on the exact position of $\tilde{u}$. This causes a 
difficulty in the use of 
(\ref{hiuv}) for the derivation of $\mo[s_i(u)]$. Nevertheless, the latter may be 
estimated with sufficient precision using the 
following facts which are true by construction/definition: 
(i) conditioned on the occurrence of the considered update
event, the probability that any other agent has changed its position
by time $\tilde{u}$ is $o(\tilde{u})$; (ii) each trajectory of $P(\cdot)$
is a continuous function; (iii) each trajectory of $P_f(\cdot)$ is a
continuous function; (iv) $\Phi$, 
the distribution function of $X_i(\tilde{u})$, is a continuous
function; (v) $g$ is a 
continuous and bounded function. Then, using the shorthand $\Xi:=
\frac{\kappa}{|J(i)|}\sum_{j\in J(i)}s_j\, -\, \left\{P-P_f\right\}$,
we have
\begin{equation}\label{hashva}
\mo\left[s_i(u)\right]= u\big\{\big(1-\Phi(-\Xi)\big)-\Phi(-\Xi)\big\}+(1-u+o(u))s_i+o(u)
\end{equation}
and consequently
\begin{equation}\label{gte}
\left(\mathcal{L}g\right)(s,P,P_f)=\big\{1-2\Phi\big(-\Xi\big)\big\}-s_i, 
\hbox{ for }g\hbox{ from }(\ref{fg}).
\end{equation}

Now let $t$ be an arbitrary real number from the interval $[0,
  t^\ast]$.  The definition~(\ref{tast}) of $t^\ast$ ensures that 
$|P(t)-P_f(t)|\leq \theta-\kappa$. This inequality and the 
obvious bound $-1\leq |J_i|^{-1}\sum_{i\in J(i)}s_{j}\leq 1$ together
  imply that
\begin{equation}\label{mejdu}
-\theta\leq \frac{\kappa}{|J_i|}\sum_{i\in J_i}s_{j}(t)-\big\{P(t)-P_f(t)\big\}
\leq \theta, \hbox{ for } t\in [0, t^\ast].
\end{equation}
By the assumption (i) of Proposition~\ref{expec}, $\Phi(x)=\frac{x+\theta}{2\theta}$ 
 when $x\in [-\theta, \theta]$. This fact combined with (\ref{mejdu})
 allows us to write the generator
\begin{eqnarray}
\big(\mathcal{L}g\big)(s(t), P(t), P_f(t))&=&
\bigg\{1-2\Phi\bigg(-\bigg[ \kappa |J(i)|^{-1}\sum_{j\in J(i)}s_j(t) - 
\left\{P(t)-P_f(t)\right\}\bigg] \bigg)\bigg\}-s_i(t)\nonumber \\
&=&-s_i(t)+\frac{\kappa}{\theta|J(i)|}\sum_{j\in J(i)}s_j(t) -
\frac{P(t)}{\theta}+\frac{P_f(t)}{\theta},\label{rol}
\end{eqnarray}
for  $t\in [0,t^\ast]$ and for $g$ from (\ref{fg}).
Combining (\ref{rol}) with (\ref{hy}), we finally get
\begin{equation}\label{pl}
\frac{d}{dt}\mo\big[s_i(t)\big]=-\mo\big[ s_i(t)\big]
+\frac{\kappa}{\theta|J_i|}\sum_{j\in J_i}\mo\big[ s_j(t)\big] 
-\frac{\mo\big[P(t)\big]}{\theta}
+\frac{\mo\big[ P_f(t)\big]}{\theta},\hbox{ for } t\in [0,t^\ast].
\end{equation}

Since $i$ was an arbitrary site of $\Lambda$, then the equation
(\ref{pl}) holds for every $i\in \Lambda$.  Hence we have a system of $\big|\Lambda\big|$ 
differential equations with the initial condition $\mo[s_i(0)]=m$,
$\forall i\in \Lambda$ (recall that, by assumption (ii) of the
proposition, the distribution of $s(0)$ has network 
homogeneous measure so that $\mo[s_i(0)]$ does not depend on $i$). It is straightforward to check 
that one of the solutions of the system of equations with this initial condition is
\begin{equation}\label{resh}
\mo[s_i(t)]=f(t),\, t\in [0, t^\ast],\,\forall i\in \Lambda,
\end{equation}
where $f(\cdot)$ is the solution of
\begin{equation}\label{foso}
\frac{d}{dt}f(t)=-\left(1-\frac{\kappa}{\theta}\right)f(t)-\frac{\mo\big[P(t)\big]}{\theta}
+\frac{\mo\big[ P_f(t)\big]}{\theta},\kern1em f(0)=m.
\end{equation} 
However, then (\ref{resh}) is the only solution, in accordance with the theory of ODEs.
Finally, (\ref{resh}) and the definition of $P(\cdot)$ imply that $P(t)=P(0)+\lambda \int_0^t f(u)du$
and (\ref{foso}) acquires the form of (\ref{difeq}). This
completes the proof of Proposition~\ref{expec}.

\subsection{Proof of Corollary 2}\label{col2p}

To prove Corollary 2 of section~\ref{regone} we shall need the
auxiliary result of Proposition 2 as stated below.  In the formulation, we use
the shorthand notation $a, b, c$ established in (\ref{short}) and
to simplify notation below we also introduce
\be
\label{tce}
t_0:=\frac{1}{2c}\log\left(\frac{a+\Delta+c}{a+\Delta-c}\right),\kern1em
\Delta:=\frac{\gamma}{\theta m(0)}.
\ee

\begin{Proposition}\label{PF}
Let $\theta>0$, $\kappa\in [0, \theta)$, $\lambda >0$, $\gamma$, $m(0), |m(0)|\leq 1$, and 
$P(0)$ be arbitrary real numbers. Let the function $f(t), t\geq 0$, 
be the solution of the ODE
\bel{odegen-a}
f^{\prime\prime}(t)+(1-\kappa/\theta)f^\prime(t)+\left(\lambda/\theta\right)f(t)=0,
\ee
with the following initial conditions
\bel{icode-a}
f(0)=m(0), \,\, \hbox{ and }\,\, f^\prime(0)=-m(0)\left(1-\kappa/\theta\right)-\gamma/\theta.
\ee
Define 
\bel{preco-a}
F(t):=P(0)+\lambda \int_0^t f(u)du, \kern1em t\geq 0,
\ee
and
\bel{tast-a}
t^\ast:=\sup\left\{t\geq 0\, :\, \left|\gamma+\lambda\int_0^\tau f(u)du\right|
\leq \theta-\kappa\,\forall \tau\leq t\right\}.
\ee

With these definitions the proposition consists of two cases.\\
\noindent{\bf (a)} If the parameter values satisfy the inequality 
$1-\kappa/\theta-2\sqrt{\lambda/\theta}>0$ 
then the extremum points of $F(\cdot)$ on the interval\footnote{We consider
                            $(0, t^\ast)$ and not $[0, t^\ast]$ because we are not 
                            interested in the function's extrema at the interval's boundary.} 
$(0, t^\ast)$ obey the following rule.
If 
\begin{eqnarray}
&&\left|\gamma\right|\leq \theta-\kappa, \label{gama-a}\\
&& m(0)\not=0,\label{mnoz}\\
&& \frac{\gamma}{\theta m(0)}>c-a,\label{estnol}\\
&&\left|\theta m(0)e^{-(a-c)t_0}\right|<\theta-\kappa,\label{casad}
\end{eqnarray}
are all satisfied, then  $t^\ast>0$ and $F(\cdot)$ has a unique extremum point on the interval
$(0, t^\ast)$; the abscissa of this point is $t_0$ of (\ref{tce}). 
If, on the contrary, at least one of the conditions (\ref{mnoz})--(\ref{casad}) fails then either
$t^\ast=0$, or $t^\ast>0$ but $F(\cdot)$ has no extrema on $(0, t^\ast)$. 
In particular, for any parameter values, the inequality
$|\gamma|<\theta-\kappa$ is sufficient for $t^\ast>0$.

\noindent{\bf (b)} If the parameter values satisfy the inequality 
$1-\kappa/\theta -2\sqrt{\lambda/\theta}<0$ then
\begin{equation}\label{pdois}
F(t)=P(0)-\gamma+e^{-at}\left\{\gamma\cos(bt)+\frac{\lambda m(0)+a\gamma}{b}\sin(bt)\right\}, t\geq 0.
\end{equation}
If $m(0)=\gamma=0$ then $F(t)=P(0)\, \forall t$. Otherwise,
the extremum points of $F(\cdot)$ on $(0, +\infty)$ are the positive 
solutions of the equation
\begin{equation}\label{tmno}
m(0)\cos(bt)-\frac{1}{b}\left[\frac{\gamma}{\theta}+am(0)\right]\sin(bt)=0,
\end{equation}
and the number of the extremum points of $F(\cdot)$ on the interval 
$(0, t^\ast)$ obeys the following rule. Let  $0\leq t_1<t_2<\ldots$ denote the
non-negative solutions of the equation (\ref{tmno}). If the inequalities 
\begin{eqnarray}
&&|\gamma|\leq \theta-\kappa,\label{sod}\\
&& e^{-at_1}\left|\gamma\cos(bt_1)+\frac{\lambda m(0)+a\gamma}{b}\sin(bt_1)\right|
\leq\theta-\kappa,\label{cosod}\\
&& e^{-at_2}\left|\gamma\cos(bt_2)+\frac{\lambda m(0)+a\gamma}{b}\sin(bt_2)\right|
\leq\theta-\kappa\label{docos}
\end{eqnarray}
are all valid then $t^\ast=\infty$ (and consequently, all the non-null $t_i$'s 
belong to $(0, t^\ast)$).
If (\ref{sod}) and (\ref{cosod}) are valid but (\ref{docos}) is invalid, 
then $t^\ast$ is finite and may be null; in this case, $t_1>0$ 
ensures that $t^\ast>0$ and that $t_1$ is the abscissa of the
unique extremum point of $F(\cdot)$ on $(0, t^\ast)$. If (\ref{sod}) 
is valid and  both (\ref{cosod}) and
(\ref{docos}) are invalid then $t^\ast$ is finite but may be null, and $F(\cdot)$ has no extrema 
on $(0,t^\ast)$. If (\ref{sod}) is invalid, then $t^\ast=0$. Again, for any parameter 
values, the inequality $|\gamma|<\theta-\kappa$ is sufficient for $t^\ast>0$.
\end{Proposition}

\medskip\noindent{\bf How Corollary~2 follows from Proposition~\ref{PF}.}\ \ 
Recall that Corollary~2 concerns the behaviour of the model if the assumptions of 
Proposition~\ref{expec} are satisfied, when the average fundamental value evaluation is fixed (i.e., 
the condition (\ref{sluca}) is satisfied), and when the initial value
$m(0)$ coincides with
$\mo[s_i(0)]\,\forall i\in \Lambda$ (all the expectations are equal because 
of the proposition's assumption about the initial distribution of the model).
Under these conditions, the definition (\ref{tast}) of $t^\ast$ and the definition 
(\ref{difeq}-\ref{inco}) of $f(\cdot)$, both from Proposition~\ref{expec}, acquire 
the forms of (\ref{tast-a}) and (\ref{odegen-a}-\ref{icode-a})
respectively from Proposition~\ref{PF}. Thus -- by the arguments of
Section~(\ref{how}) -- $P(\cdot)$, the price function considered in 
Corollary~2, coincides on the interval $[0, t^\ast]$ with the function $F(\cdot)$ 
defined and studied in Proposition~\ref{PF}. Consequently,
\begin{description}\item{(I)} The first of the three ``if" cases of  
Corollary~2 follows from Proposition~\ref{PF}(a), 
if we show that
\begin{description}\item{(I-i)} the conditions (\ref{mnoz}, \ref{estnol}, 
\ref{gama-a}, \ref{casad}) and the inequality
$1-\kappa/\theta-2\sqrt{\lambda/\theta}>0$ hold for a non-empty set of
  the model's parameter values
(because in this case, $t^\ast>0$ and $F(\cdot)$ exhibits a single bounce on $[0, t^\ast]$);
\item{(I-ii)} there is a non-empty set of model's parameter values that satisfy 
$1-\kappa/\theta-2\sqrt{\lambda/\theta}>0$, $|\gamma|<\theta-\kappa$ but do not satisfy
at least one of the conditions (\ref{gama-a}, \ref{mnoz}, \ref{estnol}, \ref{casad})
(because in this case, $t^\ast>0$ and $F(\cdot)$ is a monotonic function on $[0,t^\ast]$).
\end{description}  
\item{(II)} The third of the three ``if" cases of  Corollary~2 follows from Proposition~\ref{PF}(b), 
if we show that
\begin{description}\item{(II-i)} the conditions (\ref{sod}, \ref{cosod}, \ref{docos}) and the inequality   
$1-\kappa/\theta-2\sqrt{\lambda/\theta}<0$ hold for a non-empty set of
  model's parameter values (because $F(\cdot)$ from (\ref{pdois}) may be expressed as 
\begin{equation}\label{Fcmos}
F(t)=Ke^{-at}\sin\left(bt+\psi\right),
\end{equation}
where $K=\sqrt{(P(0))^2+(\lambda m(0)+aP(0))^2/b^2}$ and $\psi=\arcsin\left(K^{-1}P(0)\right)$,
and because $t^\ast=\infty$).
\end{description}\end{description}

The assertion (I-ii) follows straightforwardly and we now demonstrate (I-i). 
Let $\theta>0, \kappa\in [0, \theta), \lambda>0$ be such that
$1-\kappa/\theta-2\sqrt{\lambda/\theta}>0$. We take any $m(0)\in (0, \theta^{-1}(\theta-\kappa))$.
Then, the condition (\ref{mnoz}) is automatically satisfied, and the condition (\ref{casad}) 
is satisfied because $e^{-(a-c)t_0}\leq 1$. Next, we take
any $\gamma\in (0, \theta-\kappa)$. Then, the condition (\ref{gama-a}) is satisfied. Moreover, 
for any pair of chosen values of $m(0)$ and $\gamma$, the condition (\ref{estnol}) is satisfied
because $c-a<0$. This proves (I-i).

Let us now prove (II-i). The proof builds upon the following identity: 
\begin{equation}\label{pinda}
e^{-at}\left(\gamma\cos(bt)+\frac{\lambda m(0)+a\gamma}{b}\sin(bt)\right)=Me^{-at}\sin\left(bt+\nu\right),
\end{equation}
where $M=\sqrt{\gamma^2+(\lambda m(0)+a \gamma)^2/b^2}$ and 
$\nu=\arcsin\left(M^{-1}\gamma\right)$. Obviously, for a non-empty set of parameter values,
the inequality (\ref{sod}) and the inequality $M<\theta-\kappa$ may be satisfied simultaneously. 
However, since $a>0$, then the validity of the latter inequality ensures that (\ref{cosod}) 
and (\ref{docos}) hold true independently of the actual values of 
$t_1$ and $t_2$. This proves (II-i).

Hence we have shown how Corollary 2 follows from Proposition 2.\footnote{
The middle ``if'' case of Corollary 2 can be treated in a similar manner, but since
it is of no interest for the purpose of this work we omit the proof.}  The rest of
the appendix is devoted to proving Proposition 2 itself.  We will do
this in two parts.

\medskip\noindent{\bf  Proof of Proposition~\ref{PF}(a).}\ \  
The definition (\ref{tast-a}) of $t^\ast$ and the fact that $f(\cdot)$ is continuous ensure
that $t^\ast>0$ when $|\gamma|<\theta-\kappa$. This proves the last statement of item (a).
The same definition of $t^\ast$ ensures that $t^\ast=0$ when $|\gamma|>\theta-\kappa$.
So, from now on, we assume the contrary, i.e., that (\ref{gama-a}) holds. 

We recall from Corollary~1 that the validity of the inequality
$1-\kappa/\theta-2\sqrt{\lambda/\theta}>0$ implies that $f(\cdot)$, the function  that solves 
(\ref{odegen-a})--(\ref{icode-a}), is given by (\ref{sola}).  Assume first that $m(0)=0$. It then 
follows from (\ref{sola}) that either $f\equiv 0$ (when $\gamma=0$),
or the equation $f(t)=0$ has a unique solution which is $t=0$. Since $F^\prime(\cdot)=\lambda f(\cdot)$, 
then in both cases, $F(t)$ is a monotonic function on $[0, \infty)$,
  and thus, it is necessarily so
on $[0, t^\ast]$ whatever $t^\ast$ may be.

For the remainder of the proof of item (a), we consider the case where 
$m(0)\not=0$. It then follows from (\ref{sola}) and
from the inequalities $a>0$, $c>0$, $a>c$ (all of which hold true because the values of 
$\theta, \kappa, \lambda$ satisfy the proposition's assumption and the inequality 
$1-\kappa/\theta-2\sqrt{\lambda/\theta}>0$) that the equation 
$f(t)=0$ has a unique solution on $(0, \infty)$ if and only if
\begin{equation}\label{solun}
a+\Delta-c>0.
\end{equation}
This inequality is trivially equivalent to (\ref{estnol}). Thus, we 
have proved that if (\ref{estnol})
fails then $F(\cdot)$ is a monotonic function on $[0, \infty)$, and
  must be so on $[0, t^\ast]$ whatever $t^\ast$ may be.

From now on we assume that (\ref{estnol}) (or, equivalently, (\ref{solun})) holds. This assumption 
ensures that $F(\cdot)$ has a unique extremum point on $(0, +\infty)$ whose abscissa is the 
positive solution of $f(t)=0$. We denote this solution by $t_0$; from (\ref{sola}), it follows 
readily that $t_0$ satisfies (\ref{tce}).

Let us now prove that the validity of (\ref{casad}) ensures that $t_0<t^\ast$, i.e., that the unique
extremum of $F(\cdot)$ on $(0, +\infty)$ actually belongs to $(0, t^\ast)$. (Note that, by proving 
$t_0<t^\ast$, we also ensure that $t^\ast>0$ since $t_0>0$.) First of all, we observe that 
$t_0$ is the abscissa of the unique extremum point of the function $\gamma+\lambda\int_0^tf(u)du,\, 
t\in (0, \infty)$ (because the derivative of this function is $\lambda f(t)$). This observation, 
the fact that $f(\cdot)$ is a continuous function,  and the inequality 
$|\gamma|\leq \theta-\kappa$ together imply (via the definition of $t^\ast$) that  
\begin{equation}\label{daba}
t_0<t^\ast\hbox{ if and only if }
 \left|\gamma+\lambda \int_0^{t_0} f(u)du\right|\leq  \theta-\kappa.
\end{equation}
To complete the proof of item (a), we thus have to show
that the inequality in the right hand side of the implication (\ref{daba})
is equivalent to (\ref{casad}). We now proceed with the argument establishing this equivalence.
{}From the expression (\ref{sola}) for the function $f(\cdot)$ we obtain that
\begin{equation}\label{proin}
\lambda\int_0^{t_0} f(u)du=\frac{\lambda m(0)}{2c}\left\{\frac{a+\Delta -c}{a-c}\left(e^{-(a-c)t_0}-1\right)
-\frac{a+\Delta +c}{a+c}\left(e^{-(a+c)t_0}-1\right)\right\}.
\end{equation}
The right hand side of (\ref{proin}) may be significantly simplified since
\[\frac{\lambda m(0)}{2c}\left\{ \frac{a+\Delta +c}{a+c} -\frac{a+\Delta -c}{a-c}\right\}
=\frac{\lambda m(0)}{2c}\left\{-\frac{2c\Delta}{a^2-c^2}\right\}=-\theta m(0)\Delta=-\gamma,
\]
where in the last but one step we used that $a^2-c^2=\lambda/\theta$. Thus, 
\begin{eqnarray}
\gamma+\lambda \int_0^{t_0} f(u)du&=&
\frac{\lambda m(0)}{2c}\left\{\frac{a+\Delta -c}{a-c}e^{-(a-c)t_0}-\frac{a+\Delta+c}{a+c}e^{-(a+c)t_0}\right\}\nonumber\\
&=&\frac{\lambda
  m(0)}{2c}\left\{\frac{2c}{a^2-c^2}e^{-(a-c)t_0}\right\} \nonumber \\
&=&
\theta m(0)e^{-(a-c)t_0}.\label{pri}
\end{eqnarray}
Here we have also used the identity 
\begin{equation}\label{iden}
\frac{a+\Delta-c}{a-c}=\frac{a+\Delta-c}{a+c}+\frac{2c}{a^2-c^2}
\end{equation}
which follows from simple algebraic calculation. The equality (\ref{pri}) shows that the inequality (\ref{casad}) 
is equivalent to the inequality in the right hand side of the implication  (\ref{daba}). 
The proof of (a) is thus completed.

\medskip
\noindent {\bf  Proof of Proposition~\ref{PF}(b).}\ \    The definition (\ref{tast-a}) of $t^\ast$ and the fact that 
$f(\cdot)$ is continuous ensure that $t^\ast>0$ when $|\gamma|<\theta-\kappa$. 
This proves the last statement of item (b).
The same definition of $t^\ast$ ensures that $t^\ast=0$ when $|\gamma|>\theta-\kappa$.
So, from now on, we assume the contrary, i.e., that (\ref{sod}) holds. 

We recall from Corollary~1 that the validity of the inequality
$1-\kappa/\theta-2\sqrt{\lambda/\theta}<0$ implies that $f(\cdot)$, the function  that solves 
(\ref{odegen-a})--(\ref{icode-a}), is given by (\ref{solformb}).  It
then follows by integration that
\begin{equation}\label{posle}
\gamma+\lambda \int_0^{t}f(u)du=e^{-at}\left\{\gamma\cos(bt)+\frac{\lambda m(0)+a\gamma}{b}\sin(bt)\right\}.
\end{equation}
Here we have used the identities
\begin{equation}\label{diro}
\lambda
m(0)\int_0^{t}e^{-au}\left\{\cos(bu)-\frac{a}{b}\sin(bu)\right\}du =
\frac{\lambda
  m(0)}{b}e^{-at}\sin(bt)
\end{equation}
and
\begin{equation}
-\frac{\lambda \gamma}{b\theta}
 \int_0^te^{-au}\sin(bu)du =\frac{a\gamma}{b}e^{-at}\sin(bt)+\gamma
 e^{-at}\cos(bt) -\gamma \label{dare}
\end{equation}
which are straightforwardly checked (recall
$a^2+b^2=\lambda/\theta$).  From (\ref{posle}) and the definition (\ref{preco-a}) 
of $F(\cdot)$ we deduce the expression 
(\ref{pdois}) for $F(\cdot)$. From this expression, $F(t)\equiv P(0)$, when $m(0)=\gamma=0$. 
Thus, for the rest of the proof, we assume that at least one of $m(0)$ and $\gamma$ is non-zero.

Since the equation $f(t)=0$ acquires the form of (\ref{tmno}) and since 
$F^\prime(\cdot)=\lambda f(\cdot)$ then the solutions of (\ref{tmno}) are the abscissa
of the extremum points of $F(\cdot)$. This proves the second assertion of item (b).

By definition $\{t_i, i=1,2,\ldots\}$ is the set of the abscissa of the extremum points of
the function $\gamma+\lambda \int_0^tf(u)du$ on $[0, \infty)$.  We now
  turn to the question of which of the $t_i$'s lie in the interval $(0,t^\ast)$.
Assume for the present that $t_1>0$; the case $t_1=0$ will be considered later.
Using this assumption, together with the definition of $t^\ast$, 
the fact that $f(\cdot)$ is continuous, and the 
inequality $|\gamma|\leq\theta-\kappa$ (as assumed above), we find that
\begin{eqnarray}
\left|\gamma+\lambda\int_0^{t_i}f(u)du\right|\leq \theta-\kappa,\,\forall i\leq k
\hbox{ and }\left|\gamma+\lambda\int_0^{t_{k+1}}f(u)du\right|> \theta-\kappa &\Rightarrow&\label{lino}\\
\kern2em \Rightarrow t_k <t^\ast<t_{k+1}.&&\label{lina}\end{eqnarray}
Let us now assume that (\ref{cosod}) holds but (\ref{docos}) does not hold. From (\ref{posle}) this
implies that $\left|\gamma+\lambda\int_0^{t_1}f(u)du\right|\leq \theta-\kappa$ and 
$\left|\gamma+\lambda\int_0^{t_2}f(u)du\right|\leq \theta-\kappa$, or, in other words,
that (\ref{lino}) holds for $k=1$. Since
(\ref{lino})$\Rightarrow$(\ref{lina}), we then conclude that
$t_1<t^\ast<t_2$. Since $t_1>0$ then $t_1$ 
is the unique point in $(0, t^\ast)$ at which  $F(\cdot)$ attains an extremum.

Let us now prove that the validity of (\ref{sod}), (\ref{cosod}) and (\ref{docos}) ensures
that $t^\ast=\infty$. For this, we re-write (\ref{solformb}) so that $f(\cdot)$ acquires 
the following form 
\begin{equation}\label{fofi}
f(t)=Ke^{-at}\left\{\cos\left(bt+\phi\right)\right\}.
\end{equation}
Here $K= \sqrt{m(0)^2+(\gamma/(b\theta)+\gamma a m(0)/\theta)^2}$ and 
$\phi=\arccos\left\{K^{-1}m(0)\right\}$, although these expressions
are nowhere needed in our proof.
The form of (\ref{fofi}) shows that $f(t)$ is $const\,\times\,\cos(bt+\phi)$ damped by  
$e^{-at}$ with $a>0$. Combining this fact with the fact that the 
$t_i$'s are the non-negative solutions of the equation $\cos(bt+\phi)=0$ we get that the numbers
\begin{equation}\label{was}
z_i:=\int_{t_i}^{t_{i+1}}f(u)du, \kern1em i=1,2,\ldots,
\end{equation}
decrease in absolute value (i.e., $|z_i|>|z_{i+1}|$)
and form a sign-alternating sequence (i.e., $\sign(z_i)=-\sign(z_{i+1})$).
Now note that (\ref{cosod}) and (\ref{docos}) state
\begin{equation}\label{wus}\left|\gamma+\lambda\int_{0}^{t_{1}}f(u)du\right|\leq \theta-\kappa\hbox{ and } 
\left|\gamma+\lambda\int_{0}^{t_{1}}f(u)du+\lambda z_1\right|\leq \theta-\kappa.
\end{equation}
These inequalities and the properties of the sequence (\ref{was}) imply that
\begin{equation}\label{wos}
\left|\gamma+\lambda\int_{0}^{t_{1}}f(u)du+\lambda\left(z_1+z_2+\cdots+z_i\right)\right|\leq 
\theta-\kappa, \kern1em i=3,4,\ldots,
\end{equation}
or, in other words, that $|\gamma+\lambda \int_0^{t_i} f(u)du|\leq \theta-\kappa$ for $i\geq 3$.
Thus, $t^\ast>t_k,\, \forall k$, but since $t_k\uparrow \infty$ then $t^\ast=\infty$.

It only remains to consider the case $t_1=0$ in which
(\ref{cosod}) reduces to (\ref{sod}).  We recall that our proof is
already working under the assumption that (\ref{sod}) holds true.  Hence
there are only two sub-cases to be analysed: when (\ref{docos}) is valid and when it is invalid.
In the first sub-case, we argue as in the above paragraph and deduce that $t^\ast=\infty$. This 
argument works here because the sequence $\{z_i\}$ is still decreasing and sign-alternating. Actually,
the only modification necessary for the argument to work here is the substitution of 
$\lambda \int_0^{t_1}f(u)du$ by $0$. Consider now the second sub-case, i.e, assume (\ref{docos}) is 
invalid. Arguing as for the implication (\ref{lino})$\Rightarrow$(\ref{lina}), we get that
$t_1\leq t^\ast<t_2$ in this sub-case. Note that this double inequality does not 
exclude the possibility $t^\ast=0$.
It will indeed occur, if $\gamma=\theta-\kappa$ (or, $-\gamma=\theta-\kappa$) and $f(\cdot)$ is 
positive (resp., negative) in the neighbourhood of $0$. Otherwise, $t^\ast>0$. However, since $t_1=0$ then
$F(\cdot)$ has no extremum points on $(0, t^\ast)$, when $t^\ast>0$.

This completes our proof of Proposition 2 and therefore also Corollary 2.

\end{document}